\documentclass[aps,pra,superscriptaddress,twocolumn,floatfix,10pt]{revtex4-1}

\usepackage{graphicx,amsmath,amssymb,amsthm,amsfonts}
\usepackage{url}
\usepackage{float}
\usepackage{xcolor}
\usepackage{hyperref}
\usepackage{romannum}
\usepackage{lipsum}
\usepackage{bm}
\usepackage[final]{changes}
\usepackage{soul}

\def\k{{\bf k}}

\def\Eps{{\pmb{\varepsilon}}}

\def\epseff{\varepsilon_{\mathrm{eff}}}
\def\mueff{\mu_{\mathrm{eff}}}
\def\r{{\bf r}}
\def\E{{\bf E}}

\def\D{{\bf D}}
\def\H{{\bf H}}
\def\B{{\bf B}}
\def\M{{\bf M}}
\def\N{{\bf N}}
\def\L{{\bf L}}
\def\ur{{\bf\hat{e}_r}}
\def\utheta{{\bf\hat{e}_\theta}}
\def\uphi{{\bf\hat{e}_\phi}}
\def\kt{{k_\mathrm{T}}}
\def\xt{{x_\mathrm{T}}}
\def\kl{{k_\mathrm{L}}}
\def\xl{{x_\mathrm{L}}}
\def\nh{{n_\mathrm{h}}}
\def\epsh{{\epsilon_\text{h}}}
\def\eps0{{\epsilon_0}}

\definechangesauthor[name={Kevin Vynck}, color=orange]{KV}
\definechangesauthor[name=<Ashod>, color=teal]{AA}

\begin{document}

\title{Effective medium description of dense clusters of plasmonic nanoparticles with spatial dispersion}

\author{Ranjeet Dwivedi}
\affiliation{Univ. Bordeaux, CNRS, CRPP, UMR 5031, F-33600 Pessac, France}
\affiliation{ENSEMBLE3 Centre of Excellence, Wolczynska 133, 01-919 Warsaw, Poland}

\author{Ashod Aradian}
\affiliation{Univ. Bordeaux, CNRS, CRPP, UMR 5031, F-33600 Pessac, France}

\author{Virginie Ponsinet}
\affiliation{Univ. Bordeaux, CNRS, CRPP, UMR 5031, F-33600 Pessac, France}

\author{Kevin Vynck}
\affiliation{Université Claude Bernard Lyon 1, CNRS, Institut Lumière Matière (iLM), F-69622 Villeurbanne, France}

\author{Alexandre Baron}
\email{alexandre.baron@u-bordeaux.fr}
\affiliation{Univ. Bordeaux, CNRS, CRPP, UMR 5031, F-33600 Pessac, France}
\affiliation{Institut Universitaire de France, 1 rue Descartes, 75231 Paris Cedex 05, France}

\date{\today}

\begin{abstract}
\noindent We study the electromagnetic behaviour of dense, spherical clusters made of hundreds of plasmonic nanoparticules under illumination by a plane wave. Using high-precision T-matrix numerical calculations, we compute the multipolar response of clusters up to 80 nm in radius and up to 44\% in particle volume fraction. We then investigate whether it is possible to obtain an effective medium description for the clusters, taking into account weak spatial dispersion in a fully consistent way. We find that the average scattered field as well as the average inner field of the spherical cluster can be accurately reproduced by applying an extended Mie theory to an equivalent homogeneous sphere characterized by three effective parameters: an electric permittivity $\varepsilon_{\mathrm{eff}}$ and a magnetic permeability $\mu_{\mathrm{eff}}$, associated to transverse modes, and a wavevector $k_\mathrm{L}$, associated to a longitudinal mode in the sphere. Our results show that artificial magnetism arises from interparticle couplings in the dense cluster, despite inclusions not displaying any individual magnetic dipole. We also find that, although largely overlooked in the literature on metamaterials, the presence of the longitudinal mode is essential to accurately reproduce the fields of the cluster, on par with the role of artificial magnetism. Our study therefore proves that, even for high concentration in inclusions, it is possible empirically to treat a cluster of plasmonic particles as a sphere made of a spatially-dispersive homogeneous medium. This offers a practical solution facilitating the computation of electromagnetic responses of such dense random media in diverse configurations of interest for the design of metamaterials and metasurfaces.
\end{abstract}

\maketitle

\section{Introduction}

\noindent The absorption and scattering of light by a homogenous sphere excited by a plane wave are exactly described by Mie theory \cite{mie1908sattigungsstrom,bohren2008absorption}. This powerful theory consists in decomposing the exciting and scattered fields on a vector spherical harmonic basis and is widely used today to describe the absorption and scattering cross-sections of a single sphere as well as the related internal and external spatial field distributions. It is commonly used to describe light-matter interaction not only of single spheres, but also materials composed of such spheres in a variety of circumstances. In contrast, the interaction of light with inhomogenous spheres is very difficult to describe theoretically \cite{gower2021effective,yazhgur2022scattering}. Densely-packed spherical colloidal clusters of metallic or dielectric inclusions - also known as plasmonic or photonic balls - have garnered a lot of interest recently, owing to their remarkable scattering behaviors and potential applications, including non-iridescent structural coloration \cite{park2014full,vogel2015color,xiao2017bioinspired,yazhgur2022inkjet},  and Huygens meta-atoms \cite{dezert2017isotropic,dezert2019complete,elancheliyan2020tailored}. Even when the inclusions behave as small resonant electric dipoles -- say a subwavelength plasmonic particle -- the ensemble properties of the cluster are radically different to those of the inclusion because of electromagnetic interactions. Spatial correlations in the position of the inclusions, in particular, are expected to play a crucial role in dense systems \cite{vynck2023light}. Beyond being objects of mere theoretical interest, such plasmonic clusters can nowadays be manufactured following a variety of fabrication routes\cite{boal2000self,berret2011controlling,durand2011reversible, sanchez2012hydrophobic,lacava2012nanoparticle,yin2014controlled,schmitt2016formation,elancheliyan2020tailored}. These systems are of particular interest because they involve localized resonant inclusions that are assembled into a Mie resonator. This multiscale resonant nature provides a lot of leverage experimentally in the engineering of the spectral scattering characteristics of the clusters. Going further, if the clusters come in the form of a suspension or ink, they could be used to coat a surface \cite{yazhgur2022inkjet}, or even be self-assembled into a hierarchical metamaterial \cite{rockstuhl2007design}. 

For all of the above applications, it appears essential to obtain an accurate description of the electromagnetic response of the clusters. This will for example help design some desired  properties for metamaterials made of large collections of clusters, before putting any effort into actually fabricating them experimentally.

To calculate the electromagnetic properties of such complex objects, the most straightforward approach is to use purely numerical calculations: one efficient and widely used tool for this purpose is the T-matrix method~\cite{Waterman1965,Mishchenko2020}, which computes the matrix relating the components of the incident field to those of the field scattered off the object. Such ``brute force'' numerical methods are highly reliable but come with limitations.  First, calculations become highly resource-consuming when the numbers of interacting particles reaches a few hundreds inside the cluster. Second, T-matrices are intrinsically dependent on the shape and size of the computed objects, so that they need to be recalculated for every new geometrical configuration of the same physical material. Finally, in the T-matrix method, the studied object is taken as a ``black box'', transferring fields from the incoming field configuration to the outcoming one, while all processes occurring en route within the cluster are ignored. This is a strong limitation to the physical understanding of the system.

It is therefore desirable to find complementary approaches, if they are possible. A rather classical, approach consists in employing a so-called ``effective medium'' or ``homogenization'' approximation, whereby it is attempted to capture the inner composite medium's response through the use of averaged, physically meaningful parameters like the dielectric permittivity, magnetic permeability, and so on, which can then be used with Maxwell's equations to calculate both the inner and outer fields. Effective medium approaches as a means of describing complex media like dense plasmonic clusters obviously bring their lot of issues that will be discussed shortly. They do have nonetheless a strong practical interest compared to other methods like the T-matrix approach: most, if not all, full-wave numerical packages available commercially make use of effective material parameters to describe media; therefore, to obtain such parameters for dense plasmonic clusters makes sense to help promote a practical use of this kind of engineered materials by the more general scientific community, beyond the specialized nano-optics community. Furthermore, when legitimate, using effective media for field calculations drastically cuts down computational costs.

The electromagnetic homogenization of disordered assemblies of particles (also known as particulate composites) has indeed been a long-standing topic that started more than a century ago~\cite{maxwell1904colours,bruggeman1935berechnung} and has experienced many developments over the years~\cite{sihvola1999electromagnetic,mackay2015modern,belov2005homogenization,kostin1997electromagnetic,michel1995strong,agranovich2006spatial,simovski2009material,Simovski2018DispersionBook}. The properties of the effective medium are traditionally obtained with the help of so-called ``mixing laws'' which relate the relevant macroscopic parameters of the ensemble to the knowledge of the properties of the constituent materials making up the inclusions and the host material. However, in practice, the range of validity of mixing laws is very limited since their predictions quickly fail as soon as the composite system becomes complex, under the influence of one or several of the following (more or less related) circumstances: increased volume fraction in inclusions $f$ (above a few percents)~\cite{sihvola1999electromagnetic}, strongly resonant scatterers~\cite{schilder2017homogenization}, multiple scattering~\cite{mallet2005maxwell}, inter-particle couplings~\cite{Vieaud2016}, multipolar response of the inclusions~\cite{simovski2009material,blanchard2020multipolar,Simovski2018DispersionBook,guerra2022unconventional}, size distribution~\cite{Spanoudaki2001}, finite system size~\cite{Guerin2006}, and so on. To extend somewhat their range of applicability, advanced mixing laws do exist~\cite{mackay2015modern,sihvola1999electromagnetic,gower2021effective, Guerin2006}, that are able to take one or the other of the aforementioned phenomena into account, but they often require specific knowledge of the composite that is difficult to extract in realistic systems (such as precise statistics on particle positions for instance), and usually involve cumbersome calculations.

Nonetheless, the non-existence of an accurate mixing law that predicts a system's average electromagnetic parameters does not mean that it cannot be homogenized in principle. In fact, the current state of the theory does not offer a clear set of criteria specifying when a particulate system can or cannot be homogenized. Therefore, one often has to resort to empirical approaches, trying to homogeneize using various protocols and then comparing predictions to numerical, full-wave solutions of the fields.

The purpose of this article is to empirically demonstrate that an equivalent, effective medium description for clusters composed of plasmonic inclusions can be obtained indeed and that both the external (or scattered) and internal fields can be efficiently retrieved as those of an equivalent, homogeneous sphere. We do not suppose the validity of any mixing law in the process; instead, we use an appropriate generic electromagnetic formalism, namely, a generalized version of the constitutive equations for the effective medium inclusive of spatial dispersion, and we show that it is necessary for the retrieval to be successful. Remarkably, this equivalence then holds for volume fractions of inclusions as high as 0.44 in the cluster. Whether this qualifies as a true homogenization of the particulate medium will also be discussed.

To do so, we compute the Mie coefficients of the actual particle cluster with the help of high-precision numerical calculations and then carefully fit them with those of a homogeneous sphere. Though a single effective electric permittivity ($\epseff$) is sufficient to describe the scattered field at low values of $f$, we show that an additional effective magnetic permeability $\mueff$ and longitudinal wavevector quantity $\kl$ describing longitudinal modes are required to accurately describe the averaged field using Mie theory for large $f$.

\section{Plasmonic cluster definition and numerical approach}
\begin{figure}[t!]
	\begin{center}	
\includegraphics [width=\columnwidth]{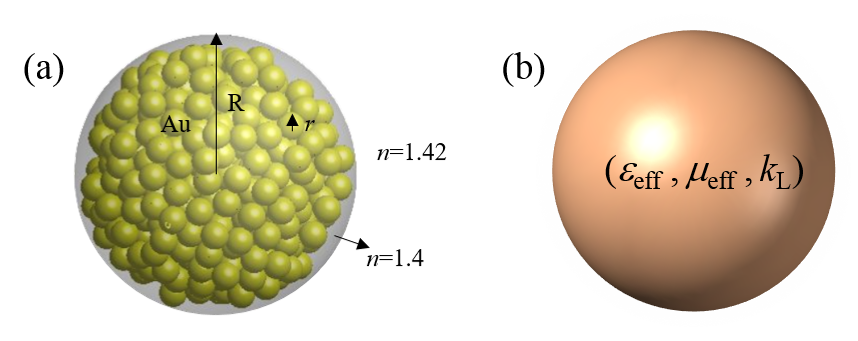}
	\end{center}
	\caption {Equivalent description of a dense plasmonic cluster. (a) The cluster of radius $R$ is composed of a dense ensemble of gold inclusions of radius $r$ embedded in a medium of refractive index 1.4. The entire cluster itself is embedded in a medium of refractive index 1.42. (b) Equivalent homogeneous sphere with effective parameters $\epseff$ alone, or $\epseff,\mueff$ and $\kl$ together, that enable the computation of the field scattered by the cluster using Mie theory.}
	\label{Fig:cluster}
\end{figure}
The system considered here is sketched on Fig.\ref{Fig:cluster}(a), it consists of an ensemble of spherical gold nanoparticles of radius $r = 7$ nm packed into a spherical cluster of radius $R$. Fixing $R$ and the volume fraction $f$ sets the number of inclusions $N$ per cluster. A rough estimate of the number of inclusions is $N\approx f(R/r)^3$. The cluster is embedded in an external medium of refractive index $n_h = 1.42$, while the inclusions within the cluster are embedded in an internal host medium of refractive index 1.4. This choice follows the experimental situation of Elancheliyan \textit{et al.}  \cite{elancheliyan2020tailored}. The dielectric function for gold is interpolated from the experimental values of Johnson and Christy~\cite{Johnson1972}. \added{We neglect corrections due to the small size of the inclusions, see Note \cite{notesmallparticles}}. The equivalent, homogeneous sphere is sketched on Fig.\ref{Fig:cluster}(b). For the purposes of our computations, the spherical cluster is numerically generated in a two-step process. First a random distribution of inclusions is generated within a cubic box of length $L$ larger than $R$ using the Lubachevsky-Stillinger algorithm \cite{lubachevsky1990geometric}. The volume fraction of inclusions is $f_b=N4\pi r^3/(3L^3)$. Then a spherical region containing all particles closer than $R-r$ to the box center is carved. The final volume fraction within this spherical cluster is thus $f=f_b(R-r)^3/R^3$. The resulting cluster formed is then simulated using the multiple sphere T-matrix software developed by Mackowski \cite{winnt,mackowski1996calculation}. The algorithm computes the multiple scattering between all inclusions up to some (high) prescribed accuracy, and returns the multipole Mie coefficients ($a_n$ and $b_n$) of the cluster for several wavelengths in the 450-1000 nm range. Since by construction, the clusters have a random inner structure of inclusions, it is necessary to repeat the entire procedure several times in order to construct a set of many cluster realizations for each given volume fraction $f$ under study. We then perform averages over each set to obtain converged averaged multipolar quantities ($\langle a_n^c\rangle,\langle b_n^c\rangle$), where the superscript \textit{c} denotes the cluster coefficients and where $\langle \cdot\rangle$ denotes ensemble averaging. This amounts to averaging the electric field scattered by the cluster, which is fully determined by the Mie coefficients. Indeed, the field averaged over $P$ random realizations, $\langle \E\rangle_P$, can be expressed as the sum of a mean coherent field $\langle \E\rangle$ and the average over $P$ realizations of a mean incoherent field $\langle \delta\E\rangle_P$ (see Appendix A):
\begin{equation}
    \langle \E\rangle_P \equiv \langle \E\rangle + \langle \delta\E\rangle_P
\end{equation}

For the composite medium to be homogenizable, one necessary condition is that $\langle \delta\E\rangle_P$, which reflects fluctuations of the field due to randomness in the system (e.g., in particle positions) vanishes as the number of realizations $P$ increases ($\lim_{P\to\infty}\langle \delta\E\rangle_P = 0$). Then, the ensemble average $\langle \E\rangle_P$ converges to some value $\langle \E\rangle$, which by definition is the field calculated in the effective medium approach. Note that this convergence is not always guaranteed, in particular for composites made of lossless resonant particles~\cite{blanchard2020multipolar,schilder2017homogenization}.

In our case involving lossy plasmonic inclusions, we do observe this condition to be fulfilled: 
we find that $P=100$ realizations are sufficient to obtain a convergence of all averaged quantities for the external field with remaining standard deviations well below one percent. 

As another interesting result from our computations, we find that, for all configurations and cluster sizes investigated, all multipoles of order $n$ larger than 2 are negligible. As a consequence, in what follows, only dipoles and quadrupoles shall be shown, with no significant loss of accuracy.

Now, to look for an equivalent effective medium, our scheme consists in fitting every retained average Mie coefficient $\langle a_{1}^c\rangle$, $\langle a_{2}^c\rangle$, $\langle b_{1}^c\rangle$, $\langle b_{2}^c\rangle$ of the cluster with those of a homogeneous sphere ($ a_{1}^h$, $ a_{2}^h$, $ b_{1}^h$, $ b_{2}^h$) of the same radius $R$, for every wavelength considered, using the Levenberg-Marquadt algorithm. As a first step, the equivalent medium in the homogeneous sphere is simply described by a local, frequency-dependent complex effective dielectric constant $\varepsilon_\mathrm{eff}(\omega)$ with its real and imaginary parts as the two fit parameters, while the effective permeability $\mu_\mathrm{eff}(\omega)=\mu_0$ is kept to the vacuum value ($\omega$ is the angular frequency). 

Let us mention that, for reference purposes only, we will be sometimes using the extended Maxwell-Garnett (MG) model in the following: this is obtained as a variant of the classical Maxwell-Garnett theory, whereby, to calculate the electric permittivity $\epseff$ and magnetic permeability $\mueff$ of the ensemble, rather than using static polarizabilities, one employs the electric and magnetic dipole coefficients (resp. $a_1$ and $b_1$) of the Mie expansion of the field scattered by individual inclusions~\cite{doyle1989optical,grimes1991permeability}.

The fitting procedure starts at high wavelengths and is initiated with a test value given by the extended MG model. Then, for every subsequent wavelength, the previously fitted solution is used as an input. The algorithm stops whenever the match between the cluster and homogeneous sphere coefficients is considered accurate enough, i.e. when $\eta = \sum_{n=1,2}\lvert \langle a_n^c\rangle -  a_n^h\rvert^2+\lvert \langle b_n^c\rangle -  b_n^h\rvert^2<10^{-4}$. 

\section{Equivalent sphere with permittivity only}

\begin{figure*}[t!]
	\centering	
\includegraphics[width=\textwidth]{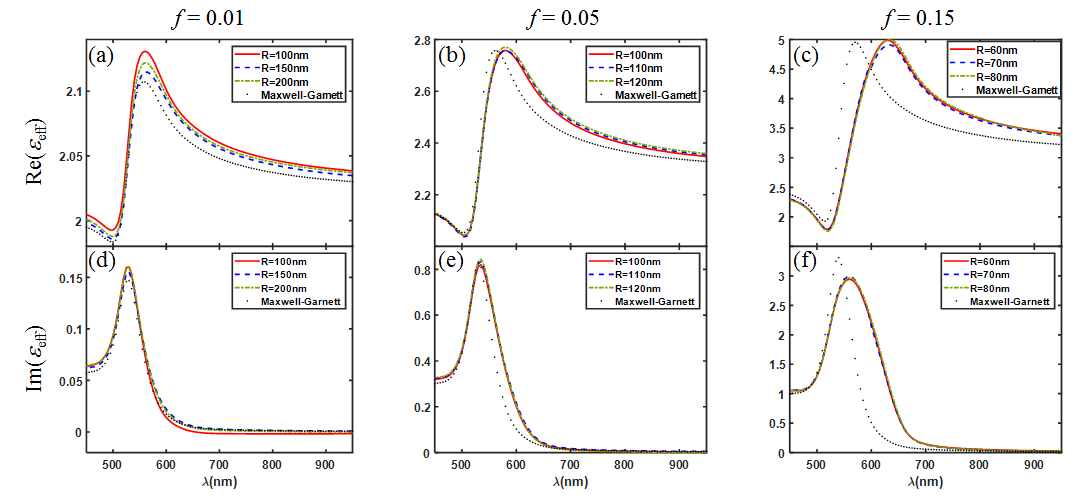}
	\caption{Spectral variations of $\varepsilon_\mathrm{eff}$ for clusters of varying volume fractions and radii. The top (bottom) panels represent the real (imaginary) parts of $\varepsilon_\mathrm{eff}$. $f$ is equal to (a,d) 0.01, (b,e) 0.05 and (c,f) 0.15. The clusters are composed of gold inclusions of radius $r=7$ nm in a sphere of radius $R$. The spectra of $\varepsilon_\mathrm{eff}$ predicted by the extended Maxwell-Garnett theory are plotted as dotted black curves for every volume fraction considered.}
	\label{Fig:smallfs}
\end{figure*}

Unsurprisingly, for dilute clusters with values of $f$ equal to 0.01, 0.05 and 0.15, we find that an effective electric permittivity $\varepsilon_\mathrm{eff}(\omega)$ is sufficient indeed to accurately fit the spectral variations of the multipole coefficients of the cluster (as shown in Fig.~\ref{Fig:smallfs} ). As should be the case, the effective effective electric permittivities we obtain (one for each $f$) are continuous functions of frequency, exhibiting a resonant behavior typical of a plasmonic system, with a gradual red-shifting of the resonance as $f$ increases. We observe that the permittivities retrieved are virtually independent of the cluster size. We note in passing that the the extended MG predictions are close to the retrieved permittivity for $f=0.01$ but that the comparison worsens quickly as $f$ increases.
\begin{figure}[b!]
	\centering	
    \includegraphics[width=0.7\columnwidth]{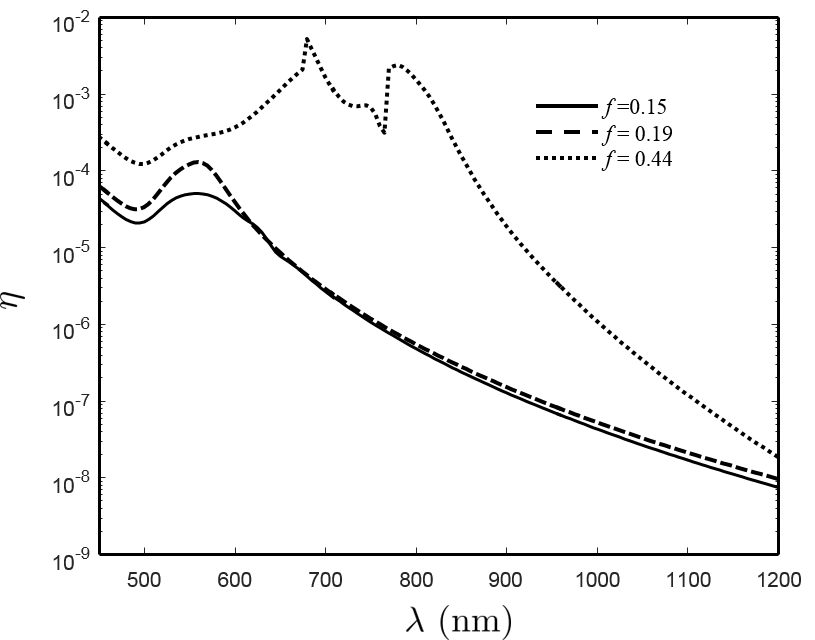}
	\caption{$\eta$ plotted as a function of wavelength for various volume fractions $f$ for a cluster of radius $R = 60$ nm.}
	\label{Fig:error}
\end{figure}

We can finally conclude that a homogeneous sphere made of a simple dielectric material featuring these $\epseff(\omega)$ permittivities is a fully satisfying equivalent for dilute plasmonic clusters \added{up to $f=0.15$. However for denser clusters, the fitting  approach fails as evidenced by Fig. \ref{Fig:error} that shows the evolution of $\eta$ when increasing $f$ for a cluster of radius $R=60$ nm. We see that the criterion $\eta<10^{-4}$ is satisfied for $f=0.15$ but starts being violated near $\lambda\approx 560$ nm when $f=0.19$. For $f=0.44$, the criterion is violated over a wide range of wavelengths from 400 nm to almost 800 nm, indicating a complete failure of the procedure.}

\begin{figure}[t!]
	\centering	
\includegraphics[width=\columnwidth]{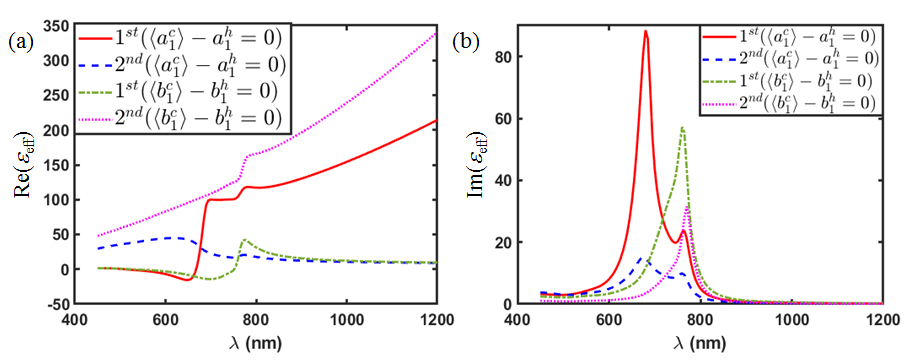}
	\caption{Solutions of $\varepsilon_\mathrm{eff}$ to the dipolar Mie coefficients for a cluster of radius $R = 60$ nm and a volume fraction $f=0.44$ of inclusions of radius $r=7$ nm. (a) Spectral variations of the (a) real and (b) imaginary parts of $\varepsilon_\mathrm{eff}$. For both $a_1$ and $b_1$, the two solutions of smallest modulus are shown.}
	\label{Fig:fail}
\end{figure}

\replaced{Therefore, we turn to this more challenging case, for which $f=0.44$ and where}{Next, we consider the more challenging case of much denser clusters for which $f=0.44$. We find that} a permittivity $\varepsilon_\mathrm{eff}$ alone is \deleted{now} insufficient to describe the scattering behavior. The reason for this is that the multipole coefficients all have resonances that cannot be simultaneously fitted with a single complex coefficient. We emphasize that this fitting failure is intrinsic, and not a mere numerical accident; it is therefore physically meaningful. To demonstrate this fact, we fit each dipolar coefficient $\langle a_{1}^c\rangle$ and $\langle b_{1}^c\rangle$ independently. We explore all values of $\varepsilon_\mathrm{eff}$ in the complex plane (restricted to positive imaginary parts), such that the homogeneous sphere has $a_1^h=\langle a_1^c\rangle$. The procedure is repeated to find the values of $\varepsilon_\mathrm{eff}$ that satisfy $b_1^h=\langle b_1^c\rangle$. For these two fits to be physically compatible, there must exist at least one common solution $\varepsilon_\mathrm{eff}(\omega)$, or equivalently $\varepsilon_\mathrm{eff}(\lambda)$ (with $\lambda$ the wavelength in vacuum) that works for both of them. While this was true for the dilute cases considered before, here however, for most of the entire spectral range considered, the two sets of solutions are disjoint. This is illustrated on Fig. \ref{Fig:fail}, where the spectral variations of the two solutions of smaller modulus for $\epseff$ are shown for a cluster of radius $R = 60$ nm. Here, two possible solutions (1 and 2) are found for each of the fitted dipole coefficients. But none of these solutions obtained from fitting $a_1$ superimposes over the entire spectral range with those obtained from fitting $b_1$. Indeed superposition is either restricted to low or high wavelengths, and is even non-existent at intermediate wavelengths (from 600 nm to 900 nm). This demonstrates that, intrinsically, no homogeneous sphere with a single permittivity (or refractive index) can account for the scattering behavior of the cluster. 

As explained in the introduction, the reason for this failure is that, as the clusters become denser, strong and intricate couplings take place between the inclusions, generating a complex response which cannot be encoded using only an effective dielectric function $\varepsilon_\mathrm{eff}(\omega)$. Several physical mechanisms are expected to take over.
In particular, when the plasmonic particles come closer together than their radius, local couplings will excite their multipolar modes. In effective medium theories, multipolarity at the level of inclusions is known to result in the emergence of so-called spatial dispersion~\cite{simovski2009material, Simovski2018DispersionBook,alu2011homogenization}, or spatial non-locality. Furthermore, interparticle couplings will also trigger collective effects, i.e. the appearance of modes that are spread over several particles: this by nature is a source of non-locality in the medium as well.

\section{Equivalent sphere with spatial dispersion}

To be able to take such phenomena into account, it is therefore necessary to seek as a next step a generalized, spatially non-local description of the material \cite{landau1984electrodynamics,agranovich2013crystal,agranovich2006spatial,Agranovich2007SpatialDispersionChapter,simovski2009material,Simovski2018DispersionBook,vinogradov2002constitutive}. A generalized effective dielectric function tensor $\Eps(\r,\r',\omega)$ is now defined such that the ensemble-averaged displacement field is non-locally related to the averaged electric field: $\langle\D(\r,\omega)\rangle \equiv \int\Eps(\omega,\r,\r')\langle\E(\r')\rangle d\r'$. From this assumption, all further steps are detailed for reference in Appendix A and can also be found in~\cite{agranovich2006spatial,Agranovich2007SpatialDispersionChapter,simovski2009material,Simovski2018DispersionBook};  we will here provide an outline of the procedure. In Fourier space, the spatially non-local permittivity transforms into a generalized permittivity $\Eps(\omega,\k)$ dependent not only on the frequency $\omega$ as before, but also on the wavevector $\k$ ($\omega$ and $\k$ being treated as independent variables). To make the problem tractable, we will assume the spatial dispersion to be weak. As a result, we expand $\Eps(\omega,\k)$ in powers of $\k$ and classically keep only terms up to second order~\cite{landau1984electrodynamics,agranovich2006spatial,simovski2009material}. Note that this is enough to take into account (up to second order in $\k$) the effect of multipolar polarizability at the level of the plasmonic inclusions, including electric and magnetic dipoles and quadrupoles~\cite{agranovich2006spatial,alu2011homogenization}, and even electric octupoles~\cite{simovski2009material}. The zeroth-order term of the $\k$-expansion is found to correspond to the standard (scalar) effective permittivity $\epseff(\omega)$ used before. First-order terms correspond to bi-anisotropy, i.e., magneto-dielectric couplings linking $\E$ and $\B$ together, responsible for e.g., chirality and girotropy~\cite{landau1984electrodynamics}. Due to the symmetries of the problem (spherical inclusions in a random arrangement, forming a spherical cluster), we can safely discard such effects in the effective medium (see Note~\cite{notechiral}).

Finally, second-order terms in $\k$ bring about two contributions~\cite{agranovich2006spatial,simovski2009material}: firstly, the emergence of artificial magnetism, providing a new electromagnetic parameter, namely the effective permeability $\mueff$; secondly, the appearance of longitudinal waves in the effective medium, characterized by a wavevector modulus $\kl$. Therefore, two types of plane waves exist in such a medium and are solutions of Helmoltz equations:
\begin{eqnarray}
    \nabla^2\E_\perp &=&-\epseff\mu_\mathrm{eff} k_0^2 \E_\perp, \label{eq:HelmT} \\ 
\nabla^2\E_\parallel &=& -\kl\!^2\E_\parallel\label{eq:HelmP},
\end{eqnarray}
where $k_0=\omega/c$, and all fields are implicitly assumed to be ensemble-averages so that brackets $\langle \cdot \rangle$ have been dropped off. The first equation applies to transverse waves (fields perpendicular to $\k$), with $\E_\perp$ the average transverse field, and with the (transverse) effective dielectric permittivity $\epseff$ and the magnetic permeability $\mu_\mathrm{eff}$ as the relevant equivalent material parameters. The second equation applies to longitudinal waves (fields parallel to \k) with $\E_\parallel$ the average longitudinal field and the parameter $\kl$ as the corresponding wavevector modulus, as introduced above. Since the longitudinal permittivity is necessarily always null (as $\D$ cannot include a longitudinal component in accordance with  Maxwell-Gauss's equation)~\cite{landau1984electrodynamics,agranovich2006spatial,agranovich2006spatial}, it cannot be used as a material parameter: this is why we use instead the wavevector modulus of the longitudinal mode $\kl$ for this purpose (see Appendix A for more details).

Although the existence of longitudinal waves associated to second-order spatial dispersion has been a long-known fact, even from classical textbooks~\cite{landau1984electrodynamics}, they have been consistently ignored in the literature devoted to the homogenization of metamaterials, resulting in the neglection of the $\kl$ parameter (or equivalent parameters), while keeping only the parameter for artificial magnetism $\mueff$. Based on rigorous theoretical derivations, this has been sternly criticized by several authors as unjustified and faulty in general \cite{agranovich2006spatial,Agranovich2007SpatialDispersionChapter,simovski2009material,Simovski2018DispersionBook}, both terms being of the same order in the $\k$-expansion. In accordance, we will seek to describe the scattering of the dense clusters by that of an equivalent homogeneous sphere carved in such a weakly spatially dispersive medium with all three parameters $\epseff$, $\mueff$ and $\kl$. To our knowledge, this work is the first attempt at retrieving in a fully consistent way this triplet of electromagnetic parameters in a plasmonic composite medium.

\begin{figure*}[t!]
	\centering	
\includegraphics[width=\textwidth]{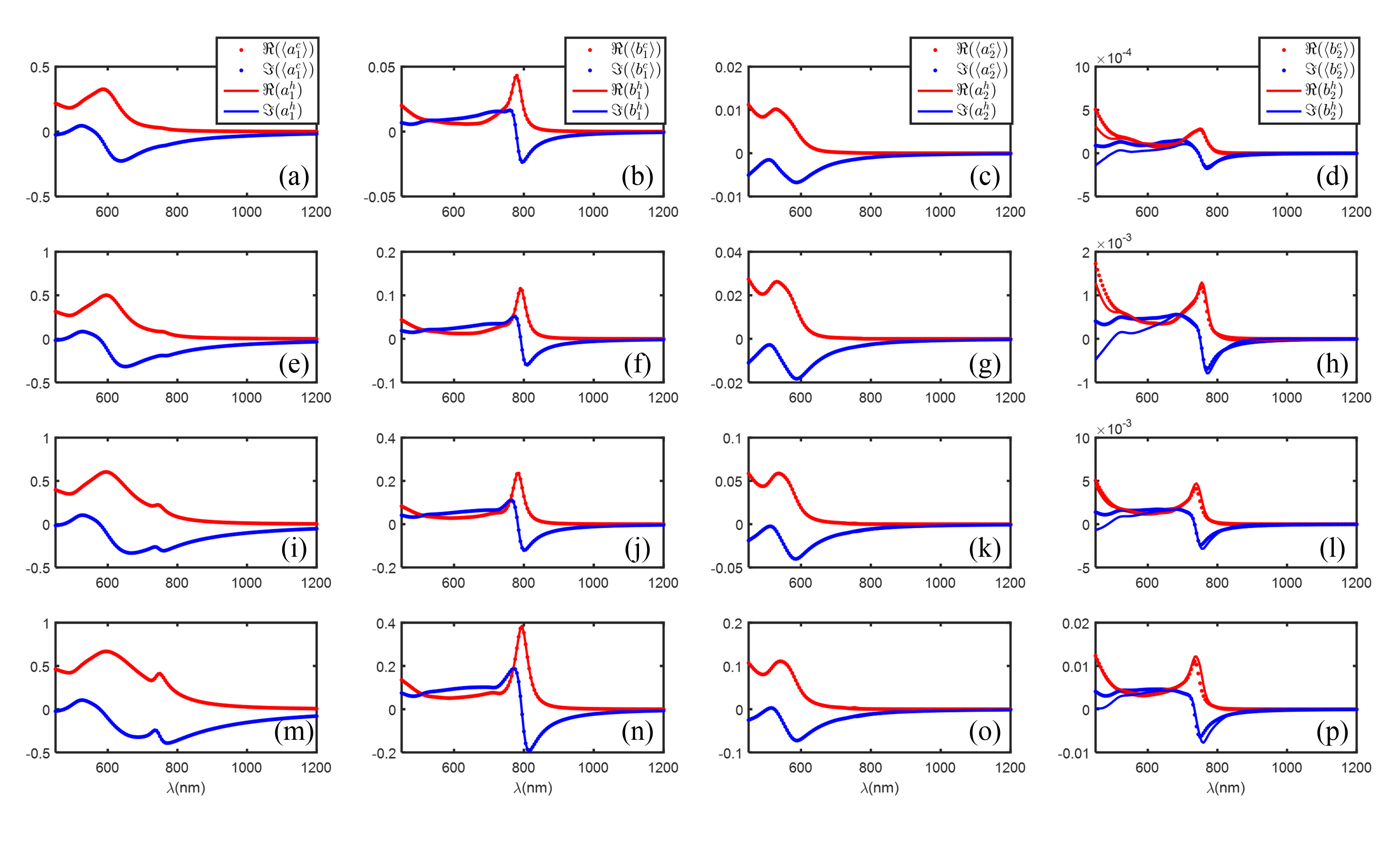}
	\caption{Comparison between the average Mie coefficients over 100 cluster realizations computed using the multiple scattering T-matrix approach and those computed from a  homogeneous sphere with the effective parameters $\varepsilon_\mathrm{eff}$, $\mu_\mathrm{eff}$, and $\kl$ retrieved from the fit. The clusters are composed of gold nanospheres of radius $r=7$ nm. The volume fill fraction is $f=0.44$. Each line corresponds to a different cluster radius: (a-d)  $R=50$ nm, (e-h)  $R=60$ nm, (i-l)  $R=70$ nm, (m-p)  $R=80$ nm. Each column shows the average of a different multipole coefficient: (a,e,i,m) $a_1$, (b,f,j,n) $b_1$, (c,g,k,o) $a_2$, (d,h,l,p) $b_2$.}
	\label{Fig:Sfits44}
\end{figure*}
Mie theory describes the scattering of light by such a material, provided some additional boundary conditions (ABCs) are given for the longitudinal waves. What the correct ABCs are for longitudinal waves in homogenous media, such as metals for instance, is a matter of debate, which lies outside our considerations. We have chosen to use the extended Mie theory provided by Ruppin \cite{ruppin1975optical}, which he originally applied to the optical properties of small metal spheres. The specifics of this theory are detailed in Appendix B. This model assumes the continuity of the normal displacement current. (Variations of this extended model for other ABCs have also been exposed~\cite{ruppin1981optical}.). The expressions we use for the $a_n$ and $b_n$ coefficients of the equivalent sphere are
\begin{widetext}
\begin{align}
a_n &= \frac {(m^2-\mu_\mathrm{eff})\kappa_n j_n(x)+j_n'(\xl)\{\mu_\mathrm{eff} \left[ mx j_n(mx)\right]'j_n(x)- m^2\left[ x j_n(x)\right]'j_n(mx)\}}{(m^2-\mu_\mathrm{eff})\kappa_n h_n(x)+j_n'(\xl)\{\mu_\mathrm{eff} \left[ mx j_n(mx)\right]'h_n(x)- m^2\left[ x h_n(x)\right]'j_n(mx)\}} \label{eq:an}\\
    b_n &= \frac{\mueff j_n(mx)\left[xj_n(x)\right]'-j_n(x)\left[mxj_n(mx)\right]'}{\mueff j_n(mx)\left[xh_n(x)\right]'-h_n(x)\left[mxj_n(mx)\right]'}\label{eq:bn}
\end{align}
\end{widetext}
where $m=(\varepsilon_\mathrm{eff}\mu_\mathrm{eff})^{1/2}/n_h$, $x=n_hk_0R$, $\xl=\kl R$ and $j_n$ and $h_n$ are respectively the spherical Bessel and Hankel functions of the first kind. Here, $\kappa_n = n(n+1)[j_n(\xl)/\xl] j_n(mx)$\added{, $n_h$ is the refractive index of the host medium, $n$ is the multipole order} and $k_0$ is the free space wave vector. Primes denote differentiation with respect to $x$. Note that the expression for $b_n$ is identical to the  classical case with no longitudinal mode~\cite{bohren2008absorption}. 

Using this three complex parameter description for the homogeneous material, we could reapply our fitting procedure to the dense clusters with $f=0.44$, for various cluster sizes. This time, we were able to find unique solutions $\epseff$, $\mueff$ and $\kl$ successfully fitting all cluster multipoles simultaneously. As can be seen on Fig. \ref{Fig:Sfits44}, the fits are excellent. Importantly, we underscore that if we tried to use only $\epseff$ and $\mueff$ for the equivalent material (thus neglecting $\kl$), we could not find any acceptable fit. 

\begin{figure*}[t!]
	\centering	
\includegraphics[width=\textwidth]{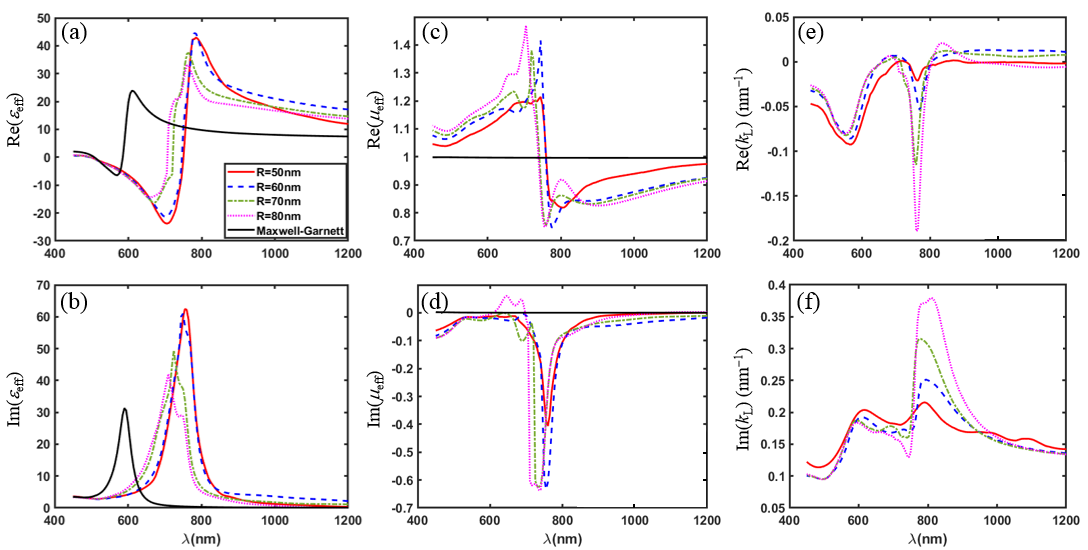}
	\caption{Spectral variations of $\varepsilon_\mathrm{eff}$, $\mu_\mathrm{eff}$ and $\kl$ for clusters varying in radius and a volume fraction of inclusions $f=0.44$. The top (bottom) panels represent the real (imaginary) parts of each parameter. The extended MG predictions are shown as black continuous lines. These predictions fail most notably because $\mu_\mathrm{eff}=1$ for the entire spectral interval since, taken separately, gold inclusions act as almost pure electric dipoles, unable to generate a magnetic activity, which in our case, results from complex interactions between particles.}
	\label{Fig:all_params}
\end{figure*}

The electric dipole and quadrupole, as well as the magnetic dipole are extremely well fitted. Only the magnetic quadrupole is poorly fitted at low wavelengths (below 600 nm). This is not a problem as it contributes very little to the scattered field and remains small compared to the other multipoles. Figure \ref{Fig:all_params} shows the spectra of all three effective parameters to which the fitting algorithm converged. All solutions are continuous. We see that $\epseff$ and $\mueff$ again exhibit a resonant behavior.

\section{Discussion: artificial magnetism and longitudinal mode}

Our results show that the homogeneous sphere has a magnetic permeability that significantly deviates from 1 even though the inclusions, taken individually, exhibit no magnetic dipole. This confirms that artificial magnetism emerges from the multiple scattering and complex couplings occurring within the cluster, as explained above. The electric permittivity reaches values as high as 40 near resonance at $\lambda = 800$ nm, corresponding to an effective index close to 6.3. These values are remarkable as there are no natural materials exhibiting such large values of the dielectric constant at optical frequencies. 

\added{We observe that $\mueff$ has a so-called ``anti-resonant'' behavior with a negative imaginary part. This type of behavior is commonplace in metamaterials (see~\cite{alu2011restoring} and references therein) and generally ascribed to the presence of spatial dispersion. It has in particular been observed in an experimental system made of clusters bearing similarities with ours~\cite{gomez2016hierarchical}. It may seem worrisome that $\mathrm{Im}(\mueff)<0$, as this may indicate that the passivity of the system is violated. It should be kept in mind, however, that all electromagnetic effects in our system arise from the electric field and its spatial derivatives solely, due to spatial dispersion; therefore this should not be interpreted as a sign that the material is active. Indeed, the artificial magnetism observed here is a consequence of spatial dispersion and as a result, the total electromagnetic power density dissipated $w\propto \varepsilon_0\mathrm{Im}(\varepsilon_\mathrm{eff})\lVert\mathbf{E}\rVert^2+\mu_0\mathrm{Im}(\mu_\mathrm{eff})\lVert\mathbf{H}\rVert^2$ should be integrated over volumes that are large enough for the spatial dispersion to produce its effect. Passivity of the system actually requires that the volume integral of $w$ remain positive, not its separate electric and magnetic terms~\cite{gomez2016hierarchical}. The overall positivity of $w$ occurs because the electrical term in the sum is positive and, when integrated over the particle volume, always dominates largely over the magnetic, negative term, in any arbitrary physically realistic field distribution. The interested reader may refere to Appendix C where we prove this assertion to hold, even in an adverse situation where it could be expected that the impact of a negative value of $\mathrm{Im}(\mueff)$ would be maximal, namely by placing the equivalent particle in a location of space where $\lVert \E\rVert \sim 0$ and $\lVert \H\rVert$ is maximal.}

Turning now to the results for the longitudinal wavevector, we observe a main peak in both $\mathrm{Re}(\kl)$ and $\mathrm{Im}(\kl)$, roughly correlated with the resonance of $\epseff$ and $\mueff$. Values for $\lvert\mathrm{Re}(\kl)\rvert$ are in the range 0--0.2, which corresponds to wavelengths $\lambda_\mathrm{L}=2\pi/\lvert\mathrm{Re}(\kl)\rvert \simeq 2\pi/0.1 \simeq 60\,$nm, i.e. typically of the same order as the size of the clusters. Looking at the penetration of this longitudinal mode, for $\lambda \leq 900$ nm (below and near resonance), the skin depth for the 80 nm cluster is equal or smaller than $\delta_\text{max}=1/[2\mathrm{Im}(\kl)] \simeq 1/(2 \times 0.1)\simeq 5\,$nm, meaning that the wave penetrates the cluster over a depth of a few particles only. Above resonance ($\lambda \geq 900$ nm), we find that $\mathrm{Im}(\kl)\gg\lvert\mathrm{Re}(\kl)\rvert$, making the longitudinal mode a purely evanescent wave.  We conclude that the longitudinal mode is essentially confined at the surface of the cluster. \added{It should be noted that both $\kl$ and its opposite are mathematically valid solutions permitting an accurate fitting of the multipoles. Therefore the choice between the two solutions is arbitrary and we decided to keep the solution for which $\mathrm{Im}(\kl)>0$.}

Importantly, we note that all parameters are dependent on the cluster size $R$, which means that the equivalence between the scattering properties of the cluster and the homogeneous sphere is numerically valid only if the equivalent parameters are used with respect to the corresponding cluster size. This comes along with the limitation that the equivalent sphere cannot be considered to be composed of a truly homogenized medium per se, for which parameters should be independent of size. This may be ascribed to the fact that our system is not large enough and remains under the influence of finite-size effects~\cite{Guerin2006}. Due to computational limitations, we were not able to check whether the equivalent parameters would converge for larger cluster sizes, and this is left as an open question of this work. In spite of this, it can be seen that both $\epseff$ and  $\mueff$ exhibit consistent trends and shapes that remain remarkably similar as $R$ increases, with changes mainly in amplitude, but not so much in spectral position. By contrast, $\kl$ displays more variability, with an increased magnitude as the radius is increased.  

\begin{figure}[t!]
	\centering	
\includegraphics[width=\columnwidth]{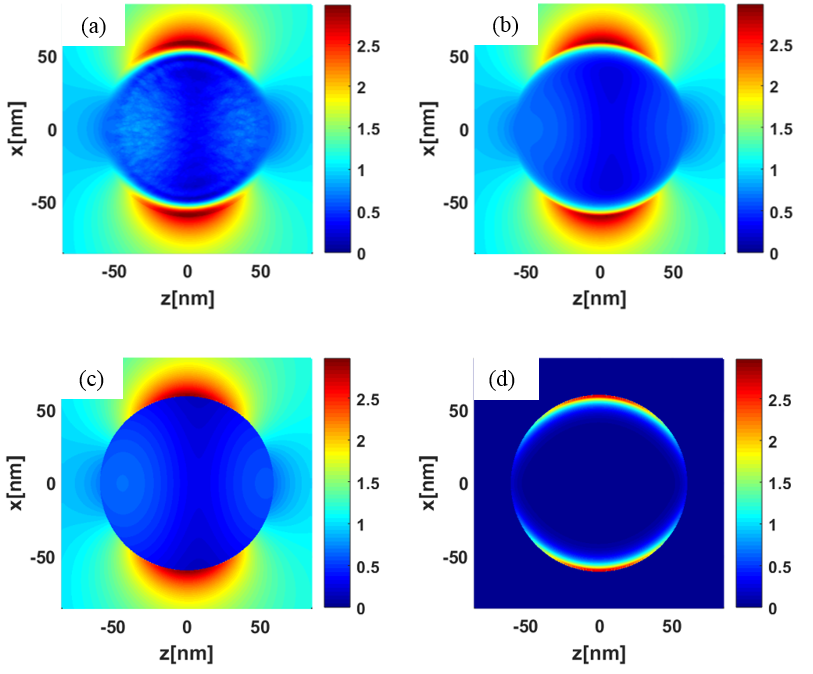}
	\caption{Comparison between the total (exciting + scattered) fields of the homogeneous, equivalent sphere and the cluster at $\lambda = 800$ nm. The cluster considered has $R=60$ nm and $f=0.44$. Plots show the magnitude of the $x$-component of the electric field, with the incoming field incident along the $z$ direction and polarized along $x$. (a) Ensemble-averaged field of the cluster. (b) Field of the homogeneous sphere. (c) Field of the transverse wave. (d) Field of the longitudinal wave.}
	\label{Fig:fields} 
\end{figure}

To further characterize our findings, on the color plots of Fig. \ref{Fig:fields}, we compare the magnitudes of the $x$-component of the average electric field of the cluster and of the electric field of the homogeneous medium for $R=60$ nm, after averaging over $P= 10^4$ realizations (the maximum number of iterations we were able to compute). In both situations the impinging plane wave has its electric field polarized along the $x$-axis and is incident along the $z$-axis. As expected, the external fields are identical, because they are composed of the exciting field and the scattered field, which was successfully fitted. We find that although the external field is very well converged to its average value after averaging over a small amount of realizations (typically $P\approx 100$), the convergence inside the cluster is much slower and many more realizations are needed to reduce the incoherent part of the internal field. \added{As a matter of fact, we observed the gradual and slow decrease of the internal average field fluctuations as a function of $P$}. As can be seen on Fig. \ref{Fig:fields}(a), even for $P=10^4$, the averaged internal field still exhibits some faint remaining incoherent fluctuations, or series of hot spots. Notwithstanding these fluctuations, we observe in Fig.~\ref{Fig:fields}(a) and (b) that the internal fields of the cluster and the equivalent sphere are strikingly similar both in symmetry and magnitude. We regard this fact as an important achievement of our effective medium retrieval method: only the external field was fitted with the help of the equivalent sphere during the procedure, yet the internal field is remarkably well reproduced as well. This brings some reassurance about the correct physical content of our findings. It furthermore hints at validating the choice of ABC made earlier as a physically plausible one, since the internal field of the homogeneous medium has been calculated from the external field through this ABC. A complete validation of this specific ABC (or other closely related ones) would however require a separate body of work on many spatially-dispersive cases.

A close inspection of Figs \ref{Fig:fields}(c) and (d) bring even more information. Figure \ref{Fig:fields}(d) illustrates the confinement of the longitudinal wave to the surface of the equivalent sphere, which shows up in the form of a corona on the plot. Figure \ref{Fig:fields}(c) on the contrary, illustrates that the transverse mode does not display such a corona. Comparing now with the actual cluster field in Fig. \ref{Fig:fields}(a), we observe the presence of a corona, which can be brought on only by the longitudinal mode. \added{This corona is not attributable to the electric or magnetic dipoles moments (see Appendix D).} This offers an indirect confirmation that the longitudinal mode we found through our fitting procedure has a physical existence.

Interestingly, the fact that a longitudinal mode would mostly play a role at the surface of the homogeneous sphere is known to other systems~\cite{LalanneBenistyGreffet2022}, including plasmonic systems such as near the boundaries of metallic nanospheres~\cite{Ciraci2013,Moreau2013}. In these systems, spatial dispersion appears due to quantum mechanical effects within the atomic layers closest to the interface. This is a different mechanism altogether to our case, but the equations with respect to the longitudinal mode are formally identical.

From the fact that the longitudinal mode is confined only to the surface of the effective medium, one could be tempted to hastily infer that its wide neglection in the literature has been legitimate. But we emphasize that it is essential to consistently transfer fields from one medium to the other: as explained above, in our study, taking it into account proved necessary for an accurate description of the cluster's scattered and internal fields, while keeping only the electric permittivity and magnetic permeability was unsuccessful.

\section{conclusion}

In conclusion, we have shown that the scattering from a dense spherical plasmonic cluster, composed of gold nanospheres that isolately act as pure electric dipoles, is well described by the scattering of a spatially dispersive homogeneous sphere with an electric permittivity, a magnetic permeability and a longitudinal wave vector. To our knowledge, our work is the first attempt at retrieving effective medium parameters for plasmonic composites, fully consistent with the theory of weakly spatially dispersive media, i.e., inclusive of both transverse and longitudinal modes. 

This set of three parameters in the extended Mie theory provides a compact formalism, which considerably reduces the complexity of describing the many-particle cluster problem while describing the scattered field with a high accuracy. Although it does not provide a full-fledged homogenization of the cluster medium, we believe that the main interest of this equivalent sphere model lies in the fact that it could be used to facilitate the prediction of the properties of metamaterials, metasurfaces or metafluids, in particular when their superstructure is composed of many such clusters: the clusters can be effectively substituted with homogeneous spheres that may be easily implemented in commercial full-wave numerical packages for instance. 

\section*{APPENDIX A: Spatial dispersion theory}
In this Appendix, we look for the expression of the constitutive equations and associated electromagnetic parameters for an \added{infinite} effective medium in presence of spatial dispersion. \added{In Appendix B, we will consider a spherical particle made of the effective medium and apply additional boundary conditions in order to extend Mie theory to the case of a finite spatially dispersive sphere.}

\subsection*{1. Generalized permittivity}

We start by expressing the average displacement vector $\langle\D(\r)\rangle$ as a function of the average electric field $\langle\E(\r)\rangle$. Averages are supposed to be taken over all possible positional configurations of the random plasmonic inclusions making up the composite medium. All fields are implicitly taken as time-harmonic functions proportional to $e^{-i\omega t}$, where $\omega$ is the angular frequency.

 In classical electromagnetism, it is assumed that the relation between $\langle\D(\r)\rangle$ and $\langle\E(\r)\rangle$ is local in space, i.e. $\langle\D(\r)\rangle = \varepsilon_0 \epseff(\omega,\r) \langle\E(\r)\rangle$. The quantity $\epseff$ is the effective permittivity and the dependence on $\omega$ is often called ``frequency dispersion''.

As explained in the main text, this is not enough to describe complex systems like dense plasmonic clusters, and one needs to consider the generalized form where the response of the medium becomes non-local, a situation usually referred to as ``spatial dispersion'' \cite{landau1984electrodynamics,agranovich2013crystal,agranovich2006spatial,Agranovich2007SpatialDispersionChapter,simovski2009material,Simovski2018DispersionBook,vinogradov2002constitutive}.

For this, we use a generalized dielectric function $\Eps(\omega,\r,\r')$, defined such that \cite{landau1984electrodynamics}
\begin{equation}
    \langle\D(\r)\rangle \equiv \varepsilon_0\int\Eps(\omega,\r,\r')\langle\E(\r')\rangle d\r'. \label{eq:SE}
\end{equation}
$\Eps(\omega,\r,\r')$ is a spatially non-local quantity that takes into account all electromagnetic interactions in a system to relate the average field $\langle\E(\r')\rangle$ at point $\r'$ to the average displacement $\langle\D(\r)\rangle$ at point $\r$. Assuming that the medium is spatially homogeneous and translationally invariant, we have $\Eps(\omega,\r,\r') = \Eps(\omega,\r-\r')$, and hence Eq. \eqref{eq:SE} becomes a convolution
\begin{equation}
      \langle\D(\r)\rangle =\varepsilon_0 \int \Eps(\omega,\r-\r') \langle\E(\r')\rangle d\r'. \label{eq:conv}
\end{equation}
In Fourier space, this last equation reduces to the following simple form
\begin{equation}
     \langle\D(\k)\rangle = \varepsilon_0 \Eps(\omega,\k) \langle\E(\k)\rangle,\label{eq:D-Sigma-E-Fourier}
\end{equation}
(Fourier transforms and their direct space counterparts are written here using the same notation).

The spatially-dispersive, generalized permittivity $\Eps(\omega,\k)$ is now a second-rank tensor, even for an isotropic system, dependent on both frequency and wavevector ($\omega$ and $\k$ being here understood as independent variables).

\subsection*{2. Weak dispersion and second-order expansion}

To make the problem tractable, we now assume that the spatial dispersion is weak, so that we can perform a Taylor expansion up to second order in $\k$ of every component of the generalized permittivity tensor $\Eps(\k) \equiv \Eps(\omega, \k)$ (from now on, we take the $\omega$-dependence as implicit):
\begin{equation}
    \varepsilon_{ij}(\mathbf{k}) \approx \varepsilon_{ij}(\mathbf{0})+\nabla_{\k}\left(\varepsilon_{ij}\right).\mathbf{k}+\frac{1}{2}\mathbf{k}^\mathrm{T}\mathcal{H}\left(\varepsilon_{ij}\right)\mathbf{k}
\end{equation}
where $\mathcal{H}\left(\varepsilon_{ij}\right)$ is the \added{$3\times3$} Hessian matrix of the $\varepsilon_{ij}$ tensor component. As explained in the main text and in Note~\cite{notechiral}, due to symmetries of the random cluster system, the averaged medium should be non-gyrotropic and non-chiral, so that the first-order term in $\k$ (known as the bi-anisotropic term) is null. 

Setting $h_{ijlm}$ as the coefficients of the fourth-rank Hessian tensor (which is built from the set of matrices $\mathcal{H}\left(\varepsilon_{ij}\right)$, when running over all indices $ij$), plugging them into the expression of $\varepsilon_{ij}$ and using Eq.~\eqref{eq:D-Sigma-E-Fourier}, we can express the expansion of each component of $\D$:
\begin{equation}
    D_{i}(\k)\approx \varepsilon_0 \varepsilon_{ij}(\mathbf{0}) E_j(\k) + \varepsilon_0 h_{ijlm} k_l k_m E_j(\k)\label{eq:Si}
\end{equation}
where $D_j$ and $E_j$ are the $j^{\mathrm{th}}$ component of $\langle\E(\k)\rangle$ and $\langle\D(\k)\rangle$, and where the implicit Einstein notation for summations has been used. 

Again due to symmetries of the problem, we assume the average medium to be isotropic, so that components $\varepsilon_{ij}$ reduce to a single value, which is the local, scalar component $\epseff$ of the generalized permittivity
\begin{equation}
    \varepsilon_{ij}(\mathbf{0}) = \epseff\delta_{ij},
\end{equation}
where $\pmb\delta$ is the Kronecker unit tensor. This $\epseff$ is the the usual effective permittivity appearing in non-spatially dispersive effective medium theories, as was successfully applied in the main text for dilute clusters.

Also following from the isotropy assumption, the fourth-rank tensor must contain only two independent parameters $(\alpha,\beta)$, and its components can be classically expressed under the form
\begin{equation}
    h_{ijlm} = \alpha \delta_{ij} \delta_{lm} + \frac{\beta}{2} \left( \delta_{il}\delta_{jm} + \delta_{im} \delta_{jl} \right).
\end{equation}
Substituting this last relation into Eq.~\eqref{eq:Si}, we find that in the end, only three coefficients are required to describe the constitutive relation of an isotropic medium with second-order spatial dispersion:
\begin{equation}
    D_i(\k) = \varepsilon_0 \epseff E_i(\k) +\varepsilon_0 \alpha k^2 E_i (\k)+ \varepsilon_0 \beta k_i k_j E_j (\k), \label{eq:Dcomp}
\end{equation}
where $k=\lVert \k \rVert$. Grouping all components $D_i$ together, we obtain the vectorial constitutive relation for $\langle \D(\k) \rangle$:
\begin{equation}
    \langle \D(\k) \rangle = \varepsilon_0 \left( \epseff + \alpha k^2 \right) \langle \E(\k) \rangle + \varepsilon_0 \beta \bigl(\k \cdot \langle \E (\k) \rangle \bigr)\, \k \label{eq:Dvector}
\end{equation}

\subsection*{3. Equivalent formalisms and artificial magnetism}

From this point on, for the sake of notational simplicity, we shall implicitly assume all fields to be averages over all configurations of the random cluster medium, i.e., $\E(\k) \equiv \langle \E(\k) \rangle$, $\D(\k) \equiv \langle \D(\k) \rangle$, etc.

With the help of the vectorial identity $\nabla\times\nabla\times\E=\nabla\,(\nabla\cdot\E) - \nabla^2\E$, or in Fourier space, $\k\times\k\times\E=(\k\cdot\E)\k-k^2\E$, the constitutive Eq.~\eqref{eq:Dvector} can be equivalently written as:
\begin{equation}
    \D(\k)=\varepsilon_0\epseff\E(\k) - \varepsilon_0\alpha\k\times\k\times\E(\k) - \varepsilon_0\gamma \bigl(\k\cdot\E(\k)\bigr) \k,
    \label{eq:Dfull-k}
\end{equation}
or, in direct space:
\begin{equation}
        \D(\r)=\varepsilon_0\epseff\E(\r) + \varepsilon_0\alpha\nabla\times\nabla\times\E(\r) + \varepsilon_0\gamma \nabla \bigl(\nabla\cdot\E(\r)\bigr),
    \label{eq:Dfull-r}
\end{equation}
where we have introduced 
\begin{equation}
\gamma = -(\alpha + \beta).
\label{eq:gammadef}
\end{equation}
Using the same vectorial identity, one can also rewrite Eq.~\eqref{eq:Dvector} in a second equivalent form:
\begin{equation}
    \D(\k)=\varepsilon_0 \left(\epseff + a k^2 \right)\E(\k) - \varepsilon_0 b\, \k\times\k\times\E(\k)
    \label{eq:DAgra}
\end{equation}
where $a=-\gamma$ and $b=-\beta$. We observe that this last equation is the same as Eq.~(23) in Agranovich and Gartstein~\cite{agranovich2006spatial}, confirming our approach to be consistent with their's.

The materials making up the composite medium have no natural magnetism ($\H=\B/\mu_0)$, therefore no microscopic currents of quantum magnetic origin exist. All currents derive from the displacement current $\partial\D / \partial t=-i\omega\D$ and are encoded in Eq.~\eqref{eq:Dfull-r}. Artificial magnetic effects will then arise from vortex-type contributions in $\D$ (current loops). Using an appropriate field transformation that keeps Maxwell's equations invariant, it can be shown~\cite{Agranovich2007SpatialDispersionChapter, agranovich2006spatial, simovski2009material,Simovski2018DispersionBook,EUMetamatBrochure} that the rotational part $\nabla\times\nabla\times\E$ in Eq.~\eqref{eq:Dfull-r} can be rigorously absorbed into an effective permeability $\mueff$, yielding the following set of equivalent material equations:
\begin{eqnarray}
    \D(\r) &=& \varepsilon_0 \epseff\E(\r)+\varepsilon_0\gamma\nabla\left(\nabla \cdot \E(\r)\right)\label{eq:D_hom}\\
    \H(\r) &=& \frac{\B(\r)}{\mu_0\mu_\mathrm{eff}\label{eq:H_hom}}
\end{eqnarray}
with
\begin{equation}
    \mu_\mathrm{eff}(\omega) = \frac{1}{1+\frac{\omega^2}{c^2}\alpha(\omega)}\:.
\end{equation}
As stated in the main text, this formulation of the material equations proves that artificial magnetism is intrinsically a second-order spatial dispersion effect, ensuing from the double curl term in Eq.~\eqref{eq:Dfull-r}. It is, however, not the only spatially-dispersive term and it should be considered on par with the term proportional to $\gamma$ in $\D$ [Eq.~\eqref{eq:Dfull-r} or \eqref{eq:D_hom}], which has been widely ignored in the metamaterials literature.

We note that, in this form, Eqs.~\eqref{eq:D_hom}-\eqref{eq:H_hom} are shown to be consistent with the approach of Simovski and Tretyakov as well, see Eqs. (2.69) and (2.70) in \cite{simovski2009material} and also \cite{EUMetamatBrochure}.

Let us now consider the consequences of formulation of Eqs.~\eqref{eq:D_hom} and \eqref{eq:H_hom}, which we will use from now on, on the propagation of waves in the effective medium.

\subsection*{4. Transverse and longitudinal wave modes}

It is useful to decompose the field in its transverse and longitudinal components, $ \E  =  \E_\perp + \E_\parallel$, with, by definition,
\begin{eqnarray}
 \k \cdot \E _\perp &=& 0, \\
 \k \times \E_\parallel &=& \mathbf{0},
\end{eqnarray}
and similarly for $\D$.
Writing Eq.~\eqref{eq:D_hom} in Fourier space:
\begin{equation}
    \D = \varepsilon_0\epseff\E - \varepsilon_0\gamma\left(\k\cdot\E\right)\k,
    \label{eq:D_hom_k}
\end{equation}
we find the expressions for the transverse and longitudinal fields, respectively,
\begin{eqnarray}
    &\D_\perp(\k) &= \varepsilon_0 \epseff \E_\perp (\k), \label{eq:D_perp} \\
    &\D_\parallel(\k) &= \varepsilon_0 \epseff \E_\parallel(\k) - \varepsilon_0 \gamma \left( \k \cdot \E_\parallel(\k)\right) \k. \label{eq:D_para}
\end{eqnarray}

Let us now investigate the propagation of plane waves inside this spatially-dispersive material. Combining Eqs. \eqref{eq:D_hom_k} and \eqref{eq:H_hom} with Maxwell's equations to find the wave equation, then separating $\E$ into transverse and parallel components, we find that these obey Helmholtz equations (as is suitable for plane waves), as follows. 

For transverse plane waves, we have
\begin{equation}
    \nabla^2\E_\perp(\r)+\epseff\mu_\mathrm{eff} k_0^2 \E_\perp(\r) = \mathbf{0}\label{eq:proptrans}
\end{equation}
with $k_0=\omega/c$. This equation gives the transverse mode dispersion relation
\begin{equation}
    \kt (\omega) = n_\mathrm{eff} (\omega) \frac{\omega}{c} = k_0 n_\mathrm{eff}(\omega),
\end{equation}
where $n_\mathrm{eff}= \pm \sqrt{\varepsilon_\mathrm{eff}\mu_\mathrm{eff}}$ is the effective index of the medium. This shows that for transverse waves, the spatially-dispersive medium acts as a medium exhibiting an effective electric permitivitty $\varepsilon_\mathrm{eff}$ and a magnetic permeability $\mu_\mathrm{eff}$.

For longitudinal plane waves, we have
\begin{equation}
    \nabla^2\E_\parallel(\r)+\kl\!^2\E_\parallel(\r) = \mathbf{0}\label{eq:proplongi},
\end{equation} 
where $\kl$ is the longitudinal mode wavevector modulus. We find the following dispersion relation for $\kl$:
\begin{equation}
    \kl(\omega) = \pm \sqrt{\frac{\epseff(\omega)}{\gamma(\omega)}} \label{eq:kldisp}
\end{equation}
This shows that the existence of the longitudinal mode in the medium is directly due to the presence of the $\gamma$-term in the material equation \eqref{eq:D_hom}.

Note that equations \eqref{eq:proptrans} and \eqref{eq:proplongi} correspond to Eqs.~\eqref{eq:HelmT} and \eqref{eq:HelmP} of the main text.

As a side note, it is possible to define from Eqs.~\ref{eq:D_perp} and \ref{eq:D_para}, transverse and longitudinal permittivities as:
\begin{eqnarray}
    &\D_\perp &= \varepsilon_0\varepsilon_\perp\E_\perp,  \\
    &\D_\parallel &= \varepsilon_0\varepsilon_\parallel \E_\parallel.
\end{eqnarray}
We find immediately that $\varepsilon_\perp= \epseff$ is the standard permittivity. After some rearrangement, we also find: $\varepsilon_\parallel =\epseff - \gamma \kl^2=0$ due to the longitudinal dispersion relation \eqref{eq:kldisp}. Therefore, as stated in the main text, the longitudinal permittivity is always null, and cannot be used as a material parameter alongside $\epseff$ and $\mueff$. This is why another parameter characterizing the longitudinal mode has to be used instead. Possible choices are explained in the next section.

\subsection*{5. Choices for effective parameters}

As the above considerations show, a medium made of a second-order spatially-dispersive material supports both transverse and longitudinal modes, and is characterized by a set of three (independent) material parameters. Several exactly equivalent choices are possible:\\
(a) From the constitutive Eq.~\eqref{eq:Dfull-r}, we may use $\epseff$, $\alpha$ and $\gamma$,\newline
(b) From the formalism of Eqs~\eqref{eq:D_hom} and \eqref{eq:H_hom}, we may use $\epseff$, $\mueff$ and $\gamma$,\newline
(c) From Eq.~\eqref{eq:kldisp}, we may use $\epseff$, $\mueff$ and $\kl$.\newline

In this work, we took choice (c), using $\mueff$ because of its physical meaning as artificial magnetism, and $\kl$ because as a wavevector modulus, it brings a more intuitive picture than the related parameter $\gamma$.

\section*{APPENDIX B: Mie theory with a spatially dispersive sphere}

We are now interested in extending Mie theory to the case of a spherical particle of electric permittivity $\epseff$, magnetic permeability $\mu_\mathrm{eff}$ and longitudinal wave vector $\kl$, immersed in a homogenous medium of dielectric constant $\epsh$. Most of the formalism presented here is adapted from Bohren and Huffmann for transverse waves \cite{bohren2008absorption}, and Ruppin for longitudinal waves \cite{ruppin1975optical}. 
We first introduce the spherical vector harmonics, in terms of which the solutions to Eqs.~\eqref{eq:proptrans} and \eqref{eq:proplongi}. There are the two following transverse functions
\begin{eqnarray}
    \M_{pmn}(\r) &=& \nabla\times(\r\psi_{pmn})\\
    \N_{pmn}(\r) &=& \frac{\nabla\times\M_{pmn}}{k}
\end{eqnarray}
and the following longitudinal function
\begin{equation}
    \L_{pmn}(\r) = \nabla(\psi_{pmn})
\end{equation}
The parity subscript stands for $e$ (even) or $o$ (odd).The generating functions are
\begin{eqnarray}
    \psi_{emn} &=& \cos(m\phi)P_n^m(\cos\theta)z_n(kr)\\
    \psi_{omn} &=& \sin(m\phi)P_n^m(\cos\theta)z_n(kr)
\end{eqnarray}
where $P_n^m(\cos\theta)$ are the associated Legendre polynimals and $z_n(kr)$ is equal to
\begin{itemize}
    \item the spherical Bessel function $j_n(\kt r)$ for transverse waves inside the sphere
    \item $j_n(\kl r)$ for longitudinal modes inside the sphere
    \item the spherical Hankel function $h_n(n_hk_0r)$ for the scattered wave outside the sphere
    \item $j_n(n_hk_0r)$ for the incident wave outside the sphere, where $k_0 = \omega/c$ is the free-space wave vector and $\nh$ is the index of refraction of the surrounding host medium in which the particles sphere is immersed.
\end{itemize}

The incident wave expands as follows using the vector spherical harmonic functions
\begin{eqnarray}
    \E_i &=& \sum_{n=1}^{\infty}E_n\left(\M_{o1n}^{(i)}-i\N_{e1n}^{(i)}\right)\\
    \H_i &=& -\frac{\nh}{c}\sum_{n=1}^{\infty}E_n\left(\M_{e1n}^{(i)}+i\N_{o1n}^{(i)}\right)
\end{eqnarray}
where $E_n = i^nE_0(2n+1)/n/(n+1)$. The scattered wave has the following expansion
\begin{eqnarray}
    \E_s &=& \sum_{n=1}^{\infty}E_n\left(ia_n\N_{e1n}^{(s)}-b_n\M_{o1n}^{(s)}\right)\\
    \H_s &=& \frac{\nh}{c}\sum_{n=1}^{\infty}E_n\left(ib_n\N_{o1n}^{(s)}+a_n\M_{e1n}^{(s)}\right)
\end{eqnarray}
The transverse wave inside the sphere is expanded as follows
\begin{eqnarray}
    \E_\mathrm{T} &=&  \sum_{n=1}^{\infty}E_n\left(c_n\M_{o1n}^{(\mathrm{T})}-id_n\N_{e1n}^{(\mathrm{T})}\right)\\
    \H_\mathrm{T} &=&  -\frac{1}{c}\sqrt{\frac{\epseff}{\mueff}}\sum_{n=1}^{\infty}E_n\left(d_n\M_{e1n}^{(\mathrm{T})}+ic_n\N_{o1n}^{(\mathrm{T})}\right)
\end{eqnarray}
Finally, the longitudinal wave inside the sphere has the following expansion
\begin{equation}
    \E_\mathrm{L} = i\sum_{n=1}^{\infty}E_nf_n\L_{e1n}
\end{equation}
and no magnetic field is associated with the longitudinal wave as a direct consequence of the fact that the longitudinal electric field is curl-free ($\nabla\times\E_\mathrm{L}=i\k\times\E_\mathrm{L}=\mathbf{0}$). 

It is useful at this stage to explicit the spherical vector functions used in these last equations
\begin{widetext}
\begin{alignat}{1}
    \M_{o1n} &= \cos\phi\pi_n(\cos\theta)z_n(kr)\utheta
    -\sin\phi\tau_n(\cos\theta)z_n(kr)\uphi\\
    \M_{e1n} &= -\sin\phi\pi_n(\cos\theta)z_n(kr)\utheta
    -\cos\phi\tau_n(\cos\theta)z_n(kr)\uphi\\
    \N_{o1n} &= \sin\phi n(n+1)\sin\theta\pi_n(\cos\theta)\frac{z_n(kr)}{kr}\ur\\
    &+ \sin\phi \tau_n(\cos\theta)\frac{\left[kr z_n(kr)\right]'}{kr}\utheta +\cos\phi\pi_n(\cos\theta)\frac{\left[kr z_n(kr)\right]'}{kr}\uphi\\
    \N_{e1n} &= \cos\phi n(n+1)\sin\theta\pi_n(\cos\theta)\frac{z_n(kr)}{kr}\ur\\
    &+ \cos\phi \tau_n(\cos\theta)\frac{\left[kr z_n(kr)\right]'}{kr}\utheta -\sin\phi\pi_n(\cos\theta)\frac{\left[kr z_n(kr)\right]'}{kr}\uphi\\
    \L_{e1n} &= \cos\phi\sin\theta\pi_n\frac{\left[kr j_n(k r)\right]'}{kr}\ur\\
    &+ \cos\phi\tau_n\frac{z_n(k r)}{r}\utheta -\sin\phi\pi_n\frac{z_n(kr)}{r}\uphi
\end{alignat}
\end{widetext}
where $\pi_n = P_n^1/\sin\theta$ and $\tau_n = dP_n^1/d\theta$.

All the Mie coefficients $a_n, b_n, c_n, d_n$ and $f_n$ are obtained from the boundary conditions at  $r=R$, where $R$ is the radius of the sphere. The tangential components of the electric and magnetic fields should be continuous, which implies
\begin{eqnarray}
    \ur\times(\E_i+\E_s)&=&\ur\times(\E_\mathrm{T}+\E_\mathrm{L})\label{eq:bound1}\\
    \ur\times(\H_i+\H_s)&=&\ur\times\H_\mathrm{T}\label{eq:bound2}
\end{eqnarray}
where $\ur$ is a unit radial vector. For media in which longitudinal polarization waves can propagate, we need an additional boundary condition (ABC) to complete this set and fully determine all the coefficients. Ruppin~\cite{ruppin1975optical} used Melnyk and Harrison's ABC~\cite{Melnyk1970Boundary}, which states that the normal displacement current should be continuous, and as a result that
\begin{equation}
    \ur.(\E_i+\E_s) = \ur.(\E_\mathrm{T}+\E_\mathrm{L})\label{eq:bound3}
\end{equation}
(the reader is referred to the main text for a brief discussion on possible ABCs).
\begin{figure*}[t!]
	\centering	
\includegraphics[width=\textwidth]{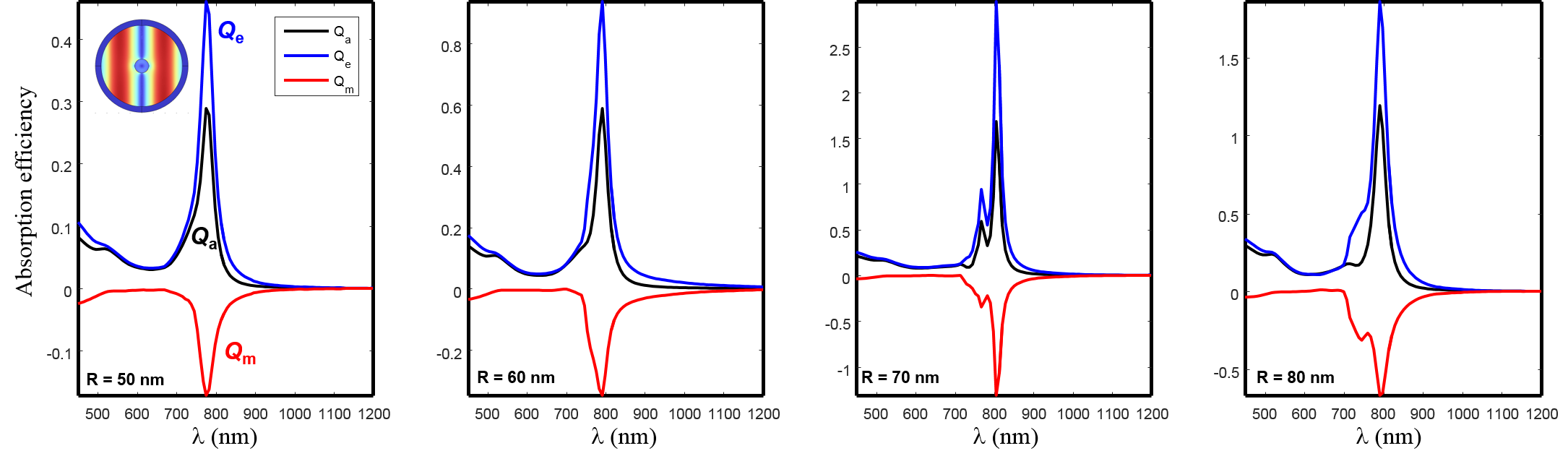}
	\caption{Total absorption $Q_a$ with electric and magnetic contributions $Q_e$ and $Q_m$ for effective particles placed in a stationary wave, with radii 50 to 80 nm. Particles are placed at a node of the electric field $\E$, which is also a maximum of field $\H$. Top-left insert: the larger sphere represents the COMSOL simulation domain with perfectly-matched layers (PMLs), the smaller sphere is the effective particle, and colors code for the magnitude of the electric field of the exciting stationary wave).}
	\label{Fig:passivity}
\end{figure*}

Exploiting eqs. \eqref{eq:bound1}, \eqref{eq:bound2} and \eqref{eq:bound3}, we find the following independent relations
\begin{widetext}
\begin{eqnarray}
    j_n(x)-b_nh_n(x)-c_nj_n(\xt)=0\\
    \left[xj_n(x)\right]'-a_n\left[xh_n(x)\right]'-\frac{d_n}{m}\left[\xt j_n(\xt)\right]'+f_nn_hk_0j_n(\xl)=0\\
    j_n(x)-a_nh_n(x)-\frac{m}{\mueff}d_nj_n(\xt)=0\\
    \left[xj_n(x)\right]'-b_n\left[xh_n(x)\right]'-\frac{c_n}{\mueff}\left[\xt j_n(\xt)\right]'=0\\
    n(n+1)j_n(x)-a_nn(n+1)h_n(x)-d_nn(n+1)\frac{j_n(\xt)}{m}+f_n\kl xj_n'(\xl)=0
\end{eqnarray}
\end{widetext}
where we have introduced the refractive index contrast $m=\kt/k_0=n_\mathrm{eff}/n_h$ and the following reduced frequencies : $x = n_hk_0R$, $\xt = \kt R$ and $\xl = \kl R$. Noticing that $\xt = mx$, we find
\begin{widetext}
\begin{eqnarray}
    a_n &=& 
\frac {(m^2-\mu_\mathrm{eff})\kappa_n j_n(x)+j_n'(\xl)\{\mu_\mathrm{eff} \left[ mx j_n(mx)\right]'j_n(x)- m^2\left[ x j_n(x)\right]'j_n(mx)\}}{(m^2-\mu_\mathrm{eff})\kappa_n h_n(x)+j_n'(\xl)\{\mu_\mathrm{eff} \left[ mx j_n(mx)\right]'h_n(x)- m^2\left[ x h_n(x)\right]'j_n(mx)\}}
\label{eq:an2}\\
    b_n &=& \frac{\mueff j_n(mx)\left[xj_n(x)\right]'-j_n(x)\left[mxj_n(mx)\right]'}{\mueff j_n(mx)\left[xh_n(x)\right]'-h_n(x)\left[mxj_n(mx)\right]'}\label{eq:bn2}
\end{eqnarray}
\end{widetext}
It should be noted that Eq. \eqref{eq:bn} is identical to that in the classical Mie theory\footnote{The interested reader can check that the equation for $b_n$ is the same as Eq. (4.53) in \cite{bohren2008absorption}.} in the absence of longitudinal modes \cite{bohren2008absorption}. So the influence of the longitudinal mode is only seen on the $a_n$ coefficient through the $\kappa_n$ coefficient and the $j_n'(\xl)$ function.

\section*{APPENDIX C: Passivity of the equivalent sphere}

\added{In this section, we evaluate the power balance of the equivalent, homogeneous particles with material parameters as shown in Fig.~\ref{Fig:all_params}, and prove that it remains dissipative. We neglect the contribution of $\kl$, since the longitudinal mode is confined to a very small region of the particle's volume, and we consider the electromagnetic power density dissipated $w = \omega/2\left(\varepsilon_0\mathrm{Im}(\varepsilon_\mathrm{eff})\lVert\mathbf{E}\rVert^2+\mu_0\mathrm{Im}(\mu_\mathrm{eff})\lVert\mathbf{H}\rVert^2\right)$. In the convention used here, $w$ is positive when the system is dissipative and negative when it is active.}

\added{To make our point, we study a ``worst-case'' scenario by placing the equivalent particles in a location of space where the exciting field satisfies $\lVert \E_\mathrm{ex}\rVert \sim 0$ and $\lVert \H_\mathrm{ex}\rVert$ is maximal: this is expected to maximize the impact of having a negative value of $\mathrm{Im}(\mueff)$. Such a situation arises by positioning the particle on an a node of a stationary electric field formed from the superposition of two counter-propagating plane waves. For particles of radius 50, 60, 70 and 80 nm, the fields were computed over the whole range of wavelengths using the finite element commercial solver Comsol Mutliphysics, using the values of $\epseff(\lambda)$ and $\mueff(\lambda)$ as shown in Fig.~\ref{Fig:all_params}. Due to the size of the particles, it is necessary to integrate the power density over their entire volume $V$ to assess whether they are dissipative or not. So we numerically compute the toal absorption cross section efficiency for each particle $Q_a = Q_e+Q_m$, which is the sum of the efficiencies of the electric ($Q_e$) and magnetic ($Q_m$) contributions defined by
\begin{eqnarray}
    Q_e &=& \frac{\omega}{2I_0\pi a^2}\int_V\varepsilon_0\mathrm{Im}(\varepsilon_\mathrm{eff})\lVert\E\rVert^2dV\\
    Q_m &=& \frac{\omega}{2I_0\pi a^2}\int_V\mu_0\mathrm{Im}(\mu_\mathrm{eff})\lVert\H\rVert^2dV
\end{eqnarray}
where $I_0$ is the intensity of the field and $a$ is the radius of the particle.   
}

\added{The result is displayed in Fig.~\ref{Fig:passivity}: it can be seen that because $\mathrm{Im}(\mueff)<0$ almost always, $Q_m<0$ also almost everywhere. However, $Q_e$ is always positive and its magnitude is much larger than $Q_m$ as well. As a result, $Q_a$ (the total absorption efficiency) is always positive in all cases (all particle radii and all wavelengths).}

\added{One may wonder how $Q_e$ can be substantially larger than $Q_m$ in a point of space where the exciting field satisfies $\lVert\E_\mathrm{ex}\rVert\sim 0$. The reason for this is that spatial dispersion extends over a region of space and therefore the effective particle can never be considered as a point-particle with respect to the wavelength $\lambda$ of the field. Because it will always extend over a region of space large enough for the spatial dispersion effects to take place, the $\E$ field is bound to vary over that domain. As soon as the value of $\E$ increases, the positive integral $Q_e$ will rise and quickly overcome the negative $Q_m$ because the imaginary part of $\epseff$ is much larger than that of $\mueff$.}

\section*{APPENDIX D: Symmetries of dipolar and corona modes}

\begin{figure}[b!]
	\centering
\includegraphics[width=\columnwidth]{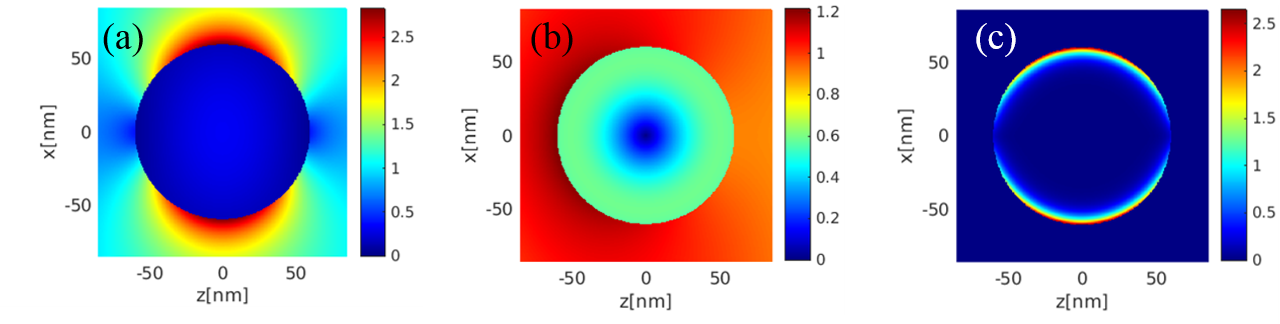}
	\caption{Color plots of the magnitude of the total electric field for the (a) electric dipole mode, (b) magnetic dipole mode and (c) longitudinal mode in the $xz-$plane of the effective medium sphere at $\lambda = 800$ nm and computed using the effective medium parameters fitted from the scattered field of the cluster medium averaged over 10$^4$ realizations. The incident field is polarized along the $x$ axis and incident along the $z$ axis.}
	\label{Fig:dipfieldplot}
\end{figure}

\added{Figure \ref{Fig:dipfieldplot} shows the magnitude of the total electric field for the electric and magnetic dipole fields. The electric dipole moment is polarized along the $x$ direction and is parallel to the polarization of the incident electric field (see Fig. \ref{Fig:dipfieldplot}(a)). The magnetic dipole moment on Fig. \ref{Fig:dipfieldplot}(b) is polarized along the $y$ axis (orthogonal to the $xz-$ plane shown on the figure). This is evidenced by a clear zero in the center of the spherical effective medium and a circulating electric field around the center. The longitudinal mode is shown on Fig. \ref{Fig:dipfieldplot}(c) and exhibits a zero along the $x=0$ line. The field is also strongly localized near the top and bottom interface. This strong localization is visible on the average field plots of the cluster medium as well, and as can be seen from the symmetries just described, it cannot be attributed to, or confused with, the electric dipole nor the magnetic dipole. Hence, the corona mode is an independent mode, associated to the longitudinal wave.}

\bibliographystyle{ieeetr}
\bibliography{library}

\begin{thebibliography}{10}

\bibitem{mie1908sattigungsstrom}
G.~Mie, ``S{\"a}ttigungsstrom und stromkurve einer schlecht leitenden fl{\"u}ssigkeit,'' {\em Annalen der Physik}, vol.~331, no.~8, pp.~597--614, 1908.

\bibitem{bohren2008absorption}
C.~F. Bohren and D.~R. Huffman, {\em Absorption and scattering of light by small particles}.
\newblock John Wiley \& Sons, 2008.

\bibitem{gower2021effective}
A.~L. Gower and G.~Kristensson, ``Effective waves for random three-dimensional particulate materials,'' {\em New Journal of Physics}, vol.~23, no.~6, p.~063083, 2021.

\bibitem{yazhgur2022scattering}
P.~Yazhgur, G.~J. Aubry, L.~S. Froufe-P{\'e}rez, and F.~Scheffold, ``Scattering phase delay and momentum transfer of light in disordered media,'' {\em Physical Review Research}, vol.~4, no.~2, p.~023235, 2022.

\bibitem{park2014full}
J.-G. Park, S.-H. Kim, S.~Magkiriadou, T.~M. Choi, Y.-S. Kim, and V.~N. Manoharan, ``Full-spectrum photonic pigments with non-iridescent structural colors through colloidal assembly,'' {\em Angewandte Chemie International Edition}, vol.~53, no.~11, pp.~2899--2903, 2014.

\bibitem{vogel2015color}
N.~Vogel, S.~Utech, G.~T. England, T.~Shirman, K.~R. Phillips, N.~Koay, I.~B. Burgess, M.~Kolle, D.~A. Weitz, and J.~Aizenberg, ``Color from hierarchy: Diverse optical properties of micron-sized spherical colloidal assemblies,'' {\em Proceedings of the National Academy of Sciences}, vol.~112, no.~35, pp.~10845--10850, 2015.

\bibitem{xiao2017bioinspired}
M.~Xiao, Z.~Hu, Z.~Wang, Y.~Li, A.~D. Tormo, N.~Le~Thomas, B.~Wang, N.~C. Gianneschi, M.~D. Shawkey, and A.~Dhinojwala, ``Bioinspired bright noniridescent photonic melanin supraballs,'' {\em Science advances}, vol.~3, no.~9, p.~e1701151, 2017.

\bibitem{yazhgur2022inkjet}
P.~Yazhgur, N.~Muller, and F.~Scheffold, ``Inkjet printing of structurally colored self-assembled colloidal aggregates,'' {\em ACS photonics}, vol.~9, no.~8, pp.~2809--2816, 2022.

\bibitem{dezert2017isotropic}
R.~Dezert, P.~Richetti, and A.~Baron, ``Isotropic huygens dipoles and multipoles with colloidal particles,'' {\em Physical Review B}, vol.~96, no.~18, p.~180201, 2017.

\bibitem{dezert2019complete}
R.~Dezert, P.~Richetti, and A.~Baron, ``Complete multipolar description of reflection and transmission across a metasurface for perfect absorption of light,'' {\em Optics express}, vol.~27, no.~19, pp.~26317--26330, 2019.

\bibitem{elancheliyan2020tailored}
R.~Elancheliyan, R.~Dezert, S.~Castano, A.~Bentaleb, E.~Nativ-Roth, O.~Regev, P.~Barois, A.~Baron, O.~Mondain-Monval, and V.~Ponsinet, ``Tailored self-assembled nanocolloidal huygens scatterers in the visible,'' {\em Nanoscale}, vol.~12, no.~47, pp.~24177--24187, 2020.

\bibitem{vynck2023light}
K.~Vynck, R.~Pierrat, R.~Carminati, L.~S. Froufe-P{\'e}rez, F.~Scheffold, R.~Sapienza, S.~Vignolini, and J.~J. S{\'a}enz, ``Light in correlated disordered media,'' {\em Reviews of Modern Physics}, vol.~95, no.~4, p.~045003, 2023.

\bibitem{boal2000self}
A.~K. Boal, F.~Ilhan, J.~E. DeRouchey, T.~Thurn-Albrecht, T.~P. Russell, and V.~M. Rotello, ``Self-assembly of nanoparticles into structured spherical and network aggregates,'' {\em Nature}, vol.~404, no.~6779, pp.~746--748, 2000.

\bibitem{berret2011controlling}
J.-F. Berret, ``Controlling electrostatic co-assembly using ion-containing copolymers: From surfactants to nanoparticles,'' {\em Advances in colloid and interface science}, vol.~167, no.~1-2, pp.~38--48, 2011.

\bibitem{durand2011reversible}
C.~Durand-Gasselin, N.~Sanson, and N.~Lequeux, ``Reversible controlled assembly of thermosensitive polymer-coated gold nanoparticles,'' {\em Langmuir}, vol.~27, no.~20, pp.~12329--12335, 2011.

\bibitem{sanchez2012hydrophobic}
A.~S{\'a}nchez-Iglesias, M.~Grzelczak, T.~Altantzis, B.~Goris, J.~Perez-Juste, S.~Bals, G.~Van~Tendeloo, S.~H. Donaldson~Jr, B.~F. Chmelka, J.~N. Israelachvili, {\em et~al.}, ``Hydrophobic interactions modulate self-assembly of nanoparticles,'' {\em ACS nano}, vol.~6, no.~12, pp.~11059--11065, 2012.

\bibitem{lacava2012nanoparticle}
J.~Lacava, P.~Born, and T.~Kraus, ``Nanoparticle clusters with lennard-jones geometries,'' {\em Nano letters}, vol.~12, no.~6, pp.~3279--3282, 2012.

\bibitem{yin2014controlled}
Q.~Yin, X.~Han, V.~Ponsinet, and H.~Liu, ``Controlled assembly of plasmonic nanoparticles using neutral-charged diblock copolymers,'' {\em Journal of colloid and interface science}, vol.~431, pp.~97--104, 2014.

\bibitem{schmitt2016formation}
J.~Schmitt, S.~Hajiw, A.~Lecchi, J.~Degrouard, A.~Salonen, M.~Imp{\'e}ror-Clerc, and B.~Pansu, ``Formation of superlattices of gold nanoparticles using ostwald ripening in emulsions: transition from fcc to bcc structure,'' {\em The Journal of Physical Chemistry B}, vol.~120, no.~25, pp.~5759--5766, 2016.

\bibitem{rockstuhl2007design}
C.~Rockstuhl, F.~Lederer, C.~Etrich, T.~Pertsch, and T.~Scharf, ``Design of an artificial three-dimensional composite metamaterial with magnetic resonances in the visible range of the electromagnetic spectrum,'' {\em Physical review letters}, vol.~99, no.~1, p.~017401, 2007.

\bibitem{Waterman1965}
P.~Waterman, ``Matrix formulation of electromagnetic scattering,'' {\em Proceedings of the IEEE}, vol.~53, no.~8, p.~805 – 812, 1965.

\bibitem{Mishchenko2020}
M.~I. Mishchenko, ``Comprehensive thematic t-matrix reference database: a 2017–2019 update,'' {\em Journal of Quantitative Spectroscopy and Radiative Transfer}, vol.~242, 2020.

\bibitem{maxwell1904colours}
J.~C. MAXWELL-GARNETT, ``Colours in metal glasses and in metallic films,'' {\em Phil. Trans. R. Soc. Lond, A}, vol.~203, pp.~385--420, 1904.

\bibitem{bruggeman1935berechnung}
V.~D. Bruggeman, ``Berechnung verschiedener physikalischer konstanten von heterogenen substanzen. i. dielektrizit{\"a}tskonstanten und leitf{\"a}higkeiten der mischk{\"o}rper aus isotropen substanzen,'' {\em Annalen der physik}, vol.~416, no.~7, pp.~636--664, 1935.

\bibitem{sihvola1999electromagnetic}
A.~H. Sihvola, {\em Electromagnetic mixing formulas and applications}.
\newblock No.~47, Iet, 1999.

\bibitem{mackay2015modern}
T.~G. Mackay and A.~Lakhtakia, {\em Modern analytical electromagnetic homogenization}.
\newblock Morgan \& Claypool Publishers San Rafael, California, USA, 2015.

\bibitem{belov2005homogenization}
P.~A. Belov and C.~R. Simovski, ``Homogenization of electromagnetic crystals formed by uniaxial resonant scatterers,'' {\em Physical Review E}, vol.~72, no.~2, p.~026615, 2005.

\bibitem{kostin1997electromagnetic}
M.~Kostin and V.~Shevchenko, ``On electromagnetic theory of artificial nonchiral and chiral media with resonant particles (a review),'' {\em Advances in Complex Electromagnetic Materials}, pp.~261--270, 1997.

\bibitem{michel1995strong}
B.~Michel and A.~Lakhtakia, ``Strong-property-fluctuation theory for homogenizing chiral particulate composites,'' {\em Physical Review E}, vol.~51, no.~6, p.~5701, 1995.

\bibitem{agranovich2006spatial}
V.~M. Agranovich and Y.~N. Gartstein, ``Spatial dispersion and negative refraction of light,'' {\em Physics-Uspekhi}, vol.~49, no.~10, p.~1029, 2006.

\bibitem{simovski2009material}
C.~R. Simovski and S.~A. Tretyakov, ``Material parameters and field energy in reciprocal composite media,'' in {\em Theory and Phenomena of Metamaterials} (F.~Capolino, ed.), CRC Press, 2009.

\bibitem{Simovski2018DispersionBook}
C.~R. Simovski, {\em Composite Media with Weak Spatial Dispersion}.
\newblock Pan Stanford Publishing, 2018.

\bibitem{schilder2017homogenization}
N.~Schilder, C.~Sauvan, Y.~R. Sortais, A.~Browaeys, and J.-J. Greffet, ``Homogenization of an ensemble of interacting resonant scatterers,'' {\em Physical Review A}, vol.~96, no.~1, p.~013825, 2017.

\bibitem{mallet2005maxwell}
P.~Mallet, C.-A. Gu{\'e}rin, and A.~Sentenac, ``Maxwell-garnett mixing rule in the presence of multiple scattering: Derivation and accuracy,'' {\em Physical Review B}, vol.~72, no.~1, p.~014205, 2005.

\bibitem{Vieaud2016}
J.~Vieaud, O.~Merchiers, M.~Rajaoarivelo, M.~Warenghem, Y.~Borensztein, V.~Ponsinet, and A.~Aradian, ``Effective medium description of plasmonic couplings in disordered polymer and gold nanoparticle composites,'' {\em Thin Solid Films}, vol.~603, p.~452 – 464, 2016.

\bibitem{blanchard2020multipolar}
C.~Blanchard, J.-P. Hugonin, A.~Nzie, and D.~D.~S. Meneses, ``Multipolar scattering of subwavelength interacting particles: Extraction of effective properties between transverse and longitudinal optical modes,'' {\em Physical Review B}, vol.~102, no.~6, p.~064209, 2020.

\bibitem{guerra2022unconventional}
T.~Guerra, D.~De~Sousa~Meneses, J.-P. Hugonin, and C.~Blanchard, ``Unconventional electromagnetic response of strongly coupled nanoparticles in the thermal infrared region: Link with effective medium properties and incoherent fields,'' {\em Particle \& Particle Systems Characterization}, vol.~39, no.~3, p.~2100245, 2022.

\bibitem{Spanoudaki2001}
A.~Spanoudaki and R.~Pelster, ``Effective dielectric properties of composite materials: The dependence on the particle size distribution,'' {\em Physical Review B - Condensed Matter and Materials Physics}, vol.~64, no.~6, 2001.

\bibitem{Guerin2006}
C.-A. Guérin, P.~Mallet, and A.~Sentenac, ``Effective-medium theory for finite-size aggregates,'' {\em Journal of the Optical Society of America A: Optics and Image Science, and Vision}, vol.~23, no.~2, p.~349 – 358, 2006.

\bibitem{Johnson1972}
P.~Johnson and R.~Christy, ``Optical constants of the noble metals,'' {\em Physical Review B}, vol.~6, no.~12, p.~4370 – 4379, 1972.

\bibitem{notesmallparticles}
Since the size of the plasmonic particles making up the cluster is comparable to the mean free path of the electrons in gold, modifications to the loss term of the bulk dielectric functions are expected. Because this is not central to the essentially theoretical purpose of our article, we neglected this effect; it could nevertheless be introduced if required, without any qualitative modification of the results.

\bibitem{lubachevsky1990geometric}
B.~D. Lubachevsky and F.~H. Stillinger, ``Geometric properties of random disk packings,'' {\em Journal of statistical Physics}, vol.~60, no.~5, pp.~561--583, 1990.

\bibitem{winnt}
D.~Mackowski, ``Mstm.'' \url{http://www.eng.auburn.edu/~dmckwski/scatcodes/}, 1999.

\bibitem{mackowski1996calculation}
D.~W. Mackowski and M.~I. Mishchenko, ``Calculation of the t matrix and the scattering matrix for ensembles of spheres,'' {\em JOSA A}, vol.~13, no.~11, pp.~2266--2278, 1996.

\bibitem{doyle1989optical}
W.~T. Doyle, ``Optical properties of a suspension of metal spheres,'' {\em Physical review B}, vol.~39, no.~14, p.~9852, 1989.

\bibitem{grimes1991permeability}
C.~A. Grimes and D.~M. Grimes, ``Permeability and permittivity spectra of granular materials,'' {\em Physical Review B}, vol.~43, no.~13, p.~10780, 1991.

\bibitem{alu2011homogenization}
A.~Al\`u, ``First-principles homogenization theory for periodic metamaterials,'' {\em Phys. Rev. B}, vol.~84, p.~075153, Aug 2011.

\bibitem{landau1984electrodynamics}
L.~Landau, E.~Lifshitz, and L.~Pitaevskii, {\em Electrodynamics of continuous media, 2nd Edition}, vol.~8.
\newblock Pergamon Press, 1984.

\bibitem{agranovich2013crystal}
V.~M. Agranovich and V.~Ginzburg, {\em Crystal optics with spatial dispersion, and excitons}, vol.~42.
\newblock Springer Science \& Business Media, 2013.

\bibitem{Agranovich2007SpatialDispersionChapter}
V.~Agranovich and G.~Y.N., ``Spatial dispersion, polaritons, and negative refraction,'' in {\em Physics of Negative Refraction and Negative Index Materials} (C.~Krowne and Y.~Zhang, eds.), Springer, 2007.

\bibitem{vinogradov2002constitutive}
A.~Vinogradov, ``On the form of constitutive equations in electrodynamics,'' {\em Physics-Uspekhi}, vol.~45, no.~3, pp.~331--338, 2002.

\bibitem{notechiral}
It is true that chiral effects, for instance, may arise in small, disordered arrays of spherical resonators; however, as soon as the number of particles increases and/or ensemble averaging is taken, inversion and mirror symmetries in the system are statistically ensured and these effects vanish (see~\cite{Pinheiro2017,Drachev2001}).

\bibitem{ruppin1975optical}
R.~Ruppin, ``Optical properties of small metal spheres,'' {\em Physical Review B}, vol.~11, no.~8, p.~2871, 1975.

\bibitem{ruppin1981optical}
R.~Ruppin, ``Optical properties of spatially dispersive dielectric spheres,'' {\em JOSA}, vol.~71, no.~6, pp.~755--758, 1981.

\bibitem{alu2011restoring}
A.~Al{\`u}, ``Restoring the physical meaning of metamaterial constitutive parameters,'' {\em Physical Review B}, vol.~83, no.~8, p.~081102, 2011.

\bibitem{gomez2016hierarchical}
S.~G{\'o}mez-Gra{\~n}a, A.~Le~Beulze, M.~Treguer-Delapierre, S.~Mornet, E.~Duguet, E.~Grana, E.~Cloutet, G.~Hadziioannou, J.~Leng, J.-B. Salmon, {\em et~al.}, ``Hierarchical self-assembly of a bulk metamaterial enables isotropic magnetic permeability at optical frequencies,'' {\em Materials Horizons}, vol.~3, no.~6, pp.~596--601, 2016.

\bibitem{LalanneBenistyGreffet2022}
H.~Benisty, J.-J. Greffet, and P.~Lalanne, {\em Introduction to Nanophotonics}.
\newblock Oxford University Press, 2022.

\bibitem{Ciraci2013}
C.~Ciracì, J.~B. Pendry, and D.~R. Smith, ``Hydrodynamic model for plasmonics: A macroscopic approach to a microscopic problem,'' {\em ChemPhysChem}, vol.~14, no.~6, p.~1109 – 1116, 2013.

\bibitem{Moreau2013}
A.~Moreau, C.~Ciracì, and D.~Smith, ``Impact of nonlocal response on metallodielectric multilayers and optical patch antennas,'' {\em Physical Review B - Condensed Matter and Materials Physics}, vol.~87, no.~4, p.~045401, 2013.

\bibitem{EUMetamatBrochure}
S.~A. Tretyakov, C.~Rockstuhl, A.~Sihvola, M.~Kafesaki, C.~Craeye, and V.~V. Kruglyak, ``Chapter 1: Basic theory of nanostructured metamaterials,'' in {\em Nanostructured Metamaterials : exchange between experts in electromagnetics and material science (Proceedings)} (V.~Kruglyak, A.~Baas, I.~Bergmair, T.~Scharf, S.~Tretyakov, and P.~Barois, eds.), EU Publications Office, 2010.
\newblock Available at \url{http://doi.org/10.2777/54953}.

\bibitem{Melnyk1970Boundary}
A.~R. Melnyk and M.~J. Harrison, ``Theory of optical excitation of plasmons in metals,'' {\em Phys. Rev. B}, vol.~2, pp.~835--850, Aug 1970.

\bibitem{Note1}
The interested reader can check that the equation for $b_n$ is the same as Eq. (4.53) in \cite {bohren2008absorption}.

\bibitem{Pinheiro2017}
F.~Pinheiro, V.~Fedotov, N.~Papasimakis, and N.~Zheludev, ``Spontaneous natural optical activity in disordered media,'' {\em Physical Review B}, vol.~95, no.~22, p.~220201, 2017.

\bibitem{Drachev2001}
V.~P. Drachev, W.~D. Bragg, V.~A. Podolskiy, V.~P. Safonov, W.-T. Kim, Z.~C. Ying, R.~L. Armstrong, and V.~M. Shalaev, ``Large local optical activity in fractal aggregates of nanoparticles,'' {\em Journal of the Optical Society of America B: Optical Physics}, vol.~18, no.~12, p.~1896 – 1903, 2001.

\end{thebibliography}

\end{document}